\documentclass[preprintnumbers,nofootinbib,showpacs,eqsecnum,pre,12pt]{revtex4-1}
\usepackage{amsmath}         
\usepackage{amssymb}          
\usepackage[pdftex]{graphicx}	
\usepackage{grffile}
\usepackage[export]{adjustbox}
\usepackage{color}  				
\usepackage[pdftex]{hyperref} 
\usepackage[utf8]{inputenc}
\usepackage{dsfont}
\usepackage{slashed}
\usepackage{array,multirow}
\usepackage{ulem}
\usepackage{xspace}

\usepackage[capitalise]{cleveref}

\crefname{section}{Sec.}{Secs.}
\crefname{table}{Tab.}{Tabs.}
\crefname{figure}{Fig.}{Figs.}
\crefname{equation}{Eq.}{Eqs.}
\crefname{appendix}{Appendix\ }{Appendices\ }

\newcommand{\ord}[1]{\mathcal{O}\left({#1}\right)}

\newcommand{\vev}{\textit{vev}\xspace}
\newcommand{\vevs}{\textit{vev}s\xspace}

\newcommand{\rsnus}{R-sneutrinos\xspace}

\newcommand{\mReSnuSq}[0]{{\ensuremath{m^{2}_{{\tilde{\nu}}^{S}}}}\xspace}
\newcommand{\mImSnuSq}[0]{{\ensuremath{m^{2}_{{\tilde{\nu}}^{P}}}}\xspace}

\newcommand{\gBL}[1]{g^{#1}_{B-L}}

\def\gsim{\raise0.3ex\hbox{$\;>$\kern-0.75em\raise-1.1ex\hbox{$\sim\;$}}}
\def\lsim{\raise0.3ex\hbox{$\;<$\kern-0.75em\raise-1.1ex\hbox{$\sim\;$}}}

\begin{document}

\preprint{XXXX, \today}
\title{Revisiting (s)neutrino dark matter in natural SUSY}
\pacs{XXXX}

 \author{T. Faber}
 \email[E-mail: ]{thomas.faber@physik.uni-wuerzburg.de}
\author{Y. Liu}
\email[E-mail: ]{yang.liu@uni-wuerzburg.de}
 \author{W. Porod}
 \email[E-mail: ]{porod@physik.uni-wuerzburg.de}
 \affiliation{
 Institut f\"ur Theoretische Physik und Astrophysik, Uni W\"urzburg
 }
 \author{J. Jones-P\'erez}
 \email[E-mail: ]{jones.j@pucp.edu.pe}
 \affiliation{
 Secci\'on F\'isica, Departamento de Ciencias, Pontificia Universidad Cat\'olica del Per\'u, Apartado 1761, Lima, Peru
 }

\begin{abstract}
We study natural supersymmetric scenarios with light right-handed neutrino superfields, and consider the possibility of having either a neutrino or a sneutrino as a dark matter candidate. For the former, we evaluate the possibility of having SUSY corrections on the $\nu_4\to\nu_\ell\gamma$ decay rate, such that the NuStar bounds are relaxed. We find that corrections are too small. For sneutrino dark matter, we consider thermal and non-thermal production, taking into account freeze-out, freeze-in and super-WIMP mechanisms. For the non-thermal case, we find that the $\tilde\nu_R$ can reproduce the observed relic density by adjusting the R-sneutrino mass and Yukawa couplings. For the thermal case, we find the need to extend the model in order to enhance sneutrino annihilations, which we exemplify in a model with an extended gauge symmetry.
\end{abstract}

\maketitle

\section{Introduction}
\label{sec:intro}

In the past decades, supersymmetric (SUSY) models have been the most popular candidates for physics beyond the Standard Model (BSM). This popularity has been very well justified, given the capacity of the Minimal Supersymmetric 
Standard Model (MSSM) of solving the hierarchy problem, providing a dark matter (DM) candidate, and achieving gauge coupling unification, among other features.

Unfortunately, in the last years there has been a lack of statistically significant signals of new physics in both collider and dark matter experiments. These constraints affect all BSM models, and in the case of SUSY they imply that either the new particles are all much heavier than expected, or that the SUSY spectrum is much more complicated than what was initially expected. This situation motivates the relaxation of assumptions typically taken in past works, such as supergravity-inspired spectra, attempting at the same time to keep most of the attractive features of such models.

In particular, we are interested in preserving naturalness as best as possible. This has important consequences, as we know from fine-tuning arguments~\cite{Papucci:2011wy} that the $\mu$ parameter should be close to the electroweak breaking scale. This implies that the lightest neutralinos $\tilde\chi^0_{1,2}$ and chargino $\tilde\chi^\pm_1$ should be of higgsino type, which leads to them being nearly mass degenerate~\cite{Barducci:2015ffa}.

Another feature that we wish to keep is having a good DM candidate. On the MSSM, the DM candidates are the lightest sneutrino $\tilde\nu_L$ and neutralino $\tilde\chi^0$, which are so-called weakly interacting massive particles (WIMPs). The $\tilde\nu_L$ has no difficulty in reproducing the correct relic density for GeV-scale masses, but in this case the spin-independent DM-nucleon cross-section lies above the current bounds, such as XENON1T~\cite{Aprile:2018dbl}. On the other hand, if the $\tilde\chi^0$ is of Higgsino type, one finds that for 
$\mathcal{O}(100~{\rm GeV})$-scale masses the correct relic density cannot be reproduced \cite{Cirelli:2005uq}. Given this situation, if we insist on having a natural SUSY solution to the DM problem, it is necessary to expand the model.

Motivated by this, we turn to one problem not addressed by the MSSM: the origin of neutrino masses. One of the most popular mechanisms addressing this issue is the Type I Seesaw~\cite{Minkowski:1977sc,Yanagida:1979as,Mohapatra:1979ia,GellMann:1980vs,Schechter:1980gr}, where new $\hat\nu_R$ superfields are added. In this case, we find ourselves with new candidates for DM: the new $\nu_R$ and $\tilde\nu_R$, provided the latter is the lightest supersymmetric particles\footnote{The scenario, where higgsino dark matter can be obtained from late sneutrino decays has
been investigated in ref.~\cite{Medina:2014bga}.}  (LSP)
\cite{Hooper:2004dc,Asaka:2005cn,Gopalakrishna:2006kr,Arina:2007tm,%
Page:2007sh,Belanger:2010cd,Dumont:2012ee,DeRomeri:2012qd,Banerjee:2016uyt,Choi:2018vdi,Ghosh:2018hly,Boyarsky:2018tvu}.

Having the $\tilde\nu_R$ as an LSP is also of phenomenological interest in the context of colliders 
\cite{deGouvea:2006wd,Thomas:2007bu,Choudhury:2008gb,%
Dumont:2012ee,Banerjee:2016uyt,Cerna-Velazco:2017cmn,Chatterjee:2017nyx,Banerjee:2018uut}. 
In ref.~\cite{Cerna-Velazco:2017cmn} it was found that there exists a 
region of the parameter space that has not been probed so far by the LHC, characterized by having a $\tilde\nu_R$ LSP, with light sleptons and higgsinos, as well as heavy gauginos. This region was also characterized by having large $Y_\nu$ couplings, which are interesting for two reasons. First, having large Yukawa couplings allows us to avoid collider bounds on long-lived sleptons. And second, this case could be probed in future experiments, such as those searching for charged lepton flavour violation, lepton number violation, or heavy neutral leptons at colliders.
It is thus of our interest to probe further this region of the parameter space, and find out what conditions do the $\nu_R$ and $\tilde\nu_R$ have to follow in order to reproduce the correct relic density. This is the main motivation behind this paper.

To this end, in \cref{sec:minimal_model}, we present the main features of a minimal model where only the $\hat\nu_R$ superfields are added. We study the conditions where thermal equilibrium can be attained, and explore thermal freeze-out as well as super-WIMP and freeze-in mechanisms. In \cref{sec:mod_BL} we add a $U(1)_{B-L}$ gauge symmetry, which is spontaneously broken by additional superfields. We find the new superfields in this model can act as mediators which help the $\tilde\nu_R$ attain thermal equilibrium.

\section{Minimal seesaw model}
\label{sec:minimal_model}

\subsection{Model definition}

As mentioned previously, we add three sterile neutrino superfields $\hat\nu_{Rk}$ ($k=4,\,5,\,6$) to the MSSM particle content, and assume conserved $R$-parity. With this, the superpotential reads as 
\begin{align}
\mathcal{W}_{eff} = \mathcal{W}_{\rm MSSM}
+ \frac{1}{2} (M_R)_{ij}\,\hat{\nu}_{Ri}\,\hat{\nu}_{Rj}
+ (Y_\nu)_{ij}\,\widehat{L}_i \cdot \widehat{H}_u\, \hat{\nu}_{Rj}
\end{align}
The corresponding soft SUSY breaking terms are given by
\begin{equation}\mathcal{V}^{soft} =\mathcal{V}_{\rm MSSM}^{soft}
  + (m^2_{\tilde\nu_R})_{ij}\tilde{\nu}^*_{Ri}\tilde{\nu}_{Rj}
  + \bigg(\frac{1}{2}(B_{\tilde\nu})_{ij}\tilde{\nu}_{Ri}\tilde{\nu}_{Rj} + (T_\nu)_{ij}\,\tilde{L}_i \cdot H_u\, \tilde{\nu}_{Rj} + \text{h.c.} \bigg)
\end{equation}

For only one family of $\hat\nu_L$/$\hat\nu_R$, the seesaw mechanism determines the size of the Yukawa couplings in terms of the neutrino masses, $Y_\nu\sim(\sqrt{2m\,M})/v_u$. If the heavy neutrino masses $M$ are on the GeV scale, one expects very small Yukawa couplings. For example, assuming a light neutrino mass $m_1=10^{-3}$~eV and $\tan\beta=6$, then, for $M=1$~GeV we have $Y_\nu=6\times10^{-9}$.

Nevertheless, when having more than one generation of $\hat\nu_R$, it is possible to enhance the Yukawas. In this case, it is useful to implement a Casa-Ibarra-like parametrization~\cite{Casas:2001sr,Donini:2012tt}. This describes neutrino mixing in terms of the active-light mixing matrix $U_{\rm PMNS}$, all six neutrino masses, and the orthogonal $R$ matrix:
\begin{equation}
 R=\left(\begin{array}{ccc}
 c_{45} & s_{45} & 0\\
 -s_{45} & c_{45} & 0 \\
 0 & 0 & 1
\end{array}\right)
\left(\begin{array}{ccc}
 c_{46} & 0 & s_{46} \\
 0 & 1 & 0 \\
 -s_{46} & 0 & c_{46} \\
\end{array}\right)
\left(\begin{array}{ccc}
1 & 0 & 0 \\
0 & c_{56} & s_{56}\\
0 & -s_{56} & c_{56}
\end{array}\right)~,
\end{equation}
where $s_{ij}$ and $c_{ij}$ are the sines and cosines of new complex
angles, $\rho_{ij}+i\gamma_{ij}$.  The $\gamma_{ij}$ lead to hyperbolic
functions, which exponentially enhance the Yukawas.

If only one of the $\gamma_{ij}$ is non-zero, the Yukawas can be expressed in a simple way. For example, by taking only $\gamma_{56}$ different from zero, 
and normal mass ordering, we find:
\begin{align}
\label{eq:Yuks}
(Y_\nu)_{a4}
= (U_{\rm PMNS})_{a1}&\sqrt{\frac{2m_1 M_4}{v_u^2}} \\
(Y_\nu)_{a5}
= -i\,z_{56}\,(Z^{\rm NO}_a)&\sqrt{\frac{2 m_3 M_5}{v_u^2}}\cosh\gamma_{56} \\
(Y_\nu)_{a6}
= (Z^{\rm NO}_a)&\sqrt{\frac{2m_3 M_6}{v_u^2}}\cosh\gamma_{56} \\
(Z^{\rm NO}_a) = (U_{\rm PMNS})_{a3}+&i\,{\rm sgn}(\gamma_{56})\,\sqrt{\frac{m_2}{m_3}}(U_{\rm PMNS})_{a1}
\end{align}
where $a=e,\,\mu,\,\tau$, and 
$m_1,\,m_2,\,m_3$ ($M_4,\,M_5,\,M_6$) are the light (heavy) neutrino 
masses. Notice that only the couplings of $\hat\nu_{R5}$ and 
$\hat\nu_{R6}$ are enhanced, with $\hat\nu_{R4}$ following the standard 
seesaw expectation. By taking $\gamma_{56}\approx3,\,5.3,\,7.6,\,9.9$, the 
elements $Y_{a5}$ and $Y_{a6}$ are enhanced by a factor 10, $10^2$, $10^3$ 
and $10^4$, respectively. Switching on the other angles does not change the generic feature that the couplings of two of the heavy neutrinos are enhanced with respect to the third one.
This is a consequence of the fact that one needs an even number of right-handed neutrinos forming pseudo-Dirac neutrinos. In the following, we shall use this scenario, setting all other $\gamma_{ij}$ and $\rho_{ij}$ to zero.
 
It is well known that when enhancing the heavy neutrino couplings one can
have large contributions to neutrinoless double beta decays from $\nu_{R5}$ and $\nu_{R6}$. The non-observation of this process forces the latter to have almost degenerate masses, 
$M_5\approx M_6$~\cite{Ibarra:2010xw,LopezPavon:2012zg,Gago:2015vma,Hernandez:2018cgc}, that is, they form a pseudo-Dirac pair. This statement holds in the presence of $R$-parity conserving SUSY, as no tree-level SUSY contributions to this process exist.

Let us now turn to the $\tilde\nu$ sector. As was discussed in~\cite{Cerna-Velazco:2017cmn}, if we neglect $B_{\tilde\nu}$ and $Y_\nu M_R^\dagger$ terms, and assume vanishing CP-violation in the sneutrino mass matrix, the real and imaginary parts of the sneutrino fields remain aligned. This means that we can work directly with complex $(\tilde\nu_L,\,\tilde\nu_R)$, with the mass matrix having the following leading terms:
\begin{equation}
 M_{\tilde\nu}^2=\left(\begin{array}{cc}
m_{\tilde L}^2+\frac{1}{2}m_Z^2\cos2\beta & 
\frac{v_u}{\sqrt{2}}\left(T_\nu-\mu Y_\nu\cot\beta\right) \\
\frac{v_u}{\sqrt{2}}\left(T^T_\nu-\mu Y^T_\nu\cot\beta\right) & 
m_{\tilde \nu_R}^2+M_R^T M_R
\end{array}\right) 
\end{equation}
Diagonalization of this mass matrix leads to the lightest mass eigenstate $\tilde\nu_1$, which in our framework is the LSP. The $\tilde\nu_1$ state will be a superposition of $\tilde\nu_R$ and $\tilde\nu_L$.
Thus, for one generation, we can have
$\tilde\nu_1=\cos\tilde\theta\,\tilde\nu_R+\sin\tilde\theta\,\tilde\nu_L$   
with L-R mixing angle:
\begin{equation}
 \label{eq:LRmixing}
 \tan2\tilde\theta\sim\frac{v_u Y_\nu}{\sqrt2}\frac{A_\nu-\mu\cot\beta}{m_{\tilde \nu_L}^2-m_{\tilde \nu_R}^2+\tfrac{1}{2}m_Z^2\cos2\beta-M_R^2}~.
\end{equation}
Here we have defined $T_\nu\equiv Y_\nu A_\nu$. The mixing is strongly suppressed by $Y_\nu$ and, given our assumptions on the size of $\mu$, can only be enhanced by taking a very large $A_\nu$, or tuning the masses such that the denominator vanishes. In the following we shall not consider any of these possibilities, such that all L-R mixing effectively vanishes.

\begin{table}
\begin{center}
\begin{tabular}{| c | c || c | c |}
\hline
Parameter & Value & Parameter & Value \\
\hline
$s_{12}^2$ & $0.310$ & $m_1$ & $10^{-3}$~eV \\
$s^2_{13}$ & $0.0224$ & $\Delta m^2_{21}$ & $7.39\times10^{-5}$~eV$^2$\\
$s^2_{23}$ & $0.582$ & $\Delta m^2_{31}$ & $2.525\times10^{-3}$~eV$^2$ \\
 &  & $M_5=M_6$ & 5~GeV \\
\hline
$\mu$ & 400~GeV & $M_1=M_2=M_3$ & $\ord{10~{\rm TeV}}$  \\
$T_\nu$ & 0~GeV & $B_\nu$ & 0~GeV$^2$ \\
& & $\tan\beta$ & 6 \\
\hline
\end{tabular}
\end{center}
\caption{\label{Table:Parameters} Fixed parameters in neutrino (top), and SUSY (bottom) sectors.}
\end{table}
In this work, for definiteness, we take the neutrino and SUSY parameters as shown in \cref{Table:Parameters}, with all CP phases equal to zero. Oscillation parameters can be found in~\cite{deSalas:2017kay,Esteban:2018azc}. $M_4$ will be allowed to vary between 1 keV and 1 MeV within our results. For the other heavy neutrino masses given in  \cref{Table:Parameters}, direct search bounds~\cite{Tanabashi:2018oca} restrict $\gamma_{56}\lesssim8$. Taking as example $\gamma_{56}=3$ ($7$), we find $|(Y_\nu)_{a5}|=|(Y_\nu)_{a6}|\sim10^{-7}$ ($10^{-5}$). On the sneutrino soft sector, only $m_{\tilde L}$ and $m_{\tilde\nu_R}$ are non-zero. Given the strong flavour constraints coming from processes such as $\mu\to e\gamma$, we take the soft masses flavour diagonal. This implies that all mixing effects are negligible, allowing us to identify the mass eigenstates with the interaction eigenstates. In particular, $\tilde\nu_1$ will be one of the three $\tilde\nu_R$ states.

\subsection{Neutrino Dark matter}
\label{sec:nuDM}

It is well known that a sterile neutrino with mass in the keV
range can serve as potential DM candidate via oscillations with
the left-handed neutrinos \cite{Dodelson:1993je,Shi:1998km}. However,
it is unstable, as it can decay into the lighter active neutrinos through the following processes:
\begin{align}
\nu_4 &\to \nu_i \gamma \\
 & \to  \nu_i \nu_j \nu_k \hspace{1cm} (i,j,k=1,\dots,3)
\end{align}
The first final state can be detected via the resulting photon,
e.g.\ by cosmological observations. In particular, data from
the NuStar collaboration puts severe constraints on the allowed 
parameter space, as has been shown recently in \cite{Ng:2019gch}.
From their Fig.~5, considering only the usual $W$-$l$-loop mediated contribution, one gets an upper bound for the quantity
\begin{equation}
  \label{eq:sin2theta}
  \sin^22\theta=\sum_{a=1}^3|U_{a4}|^2
\end{equation}
of few times $10^{-11}$ for $M_4=7$~keV and up to around $10^{-14}$ for $M_4=50$~keV. We find that the $\sin^22\theta$ predicted 
in our scenario is always well above 
the corresponding limit, as can be seen from the values given in
\cref{tab:mixcompare}  (see also~\cite{Drewes:2019mhg} for a recent discussion on the seesaw prediction).

\begin{figure}[t]
  \begin{center}
    \includegraphics[width=0.8\textwidth]{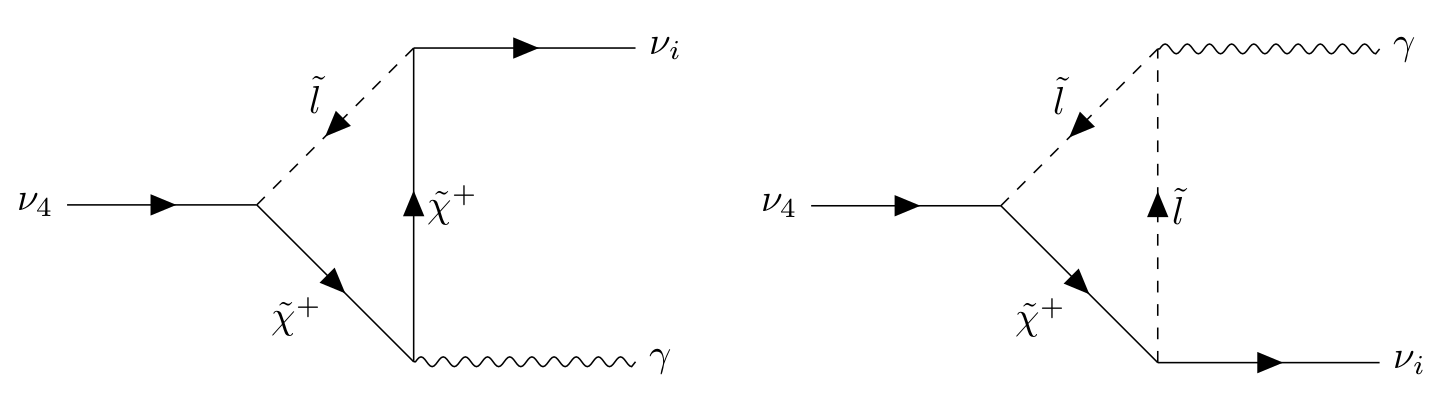}
  \end{center}
  \caption{Contributions to $\nu_4\to\nu_i\,\gamma$ from $\tilde{\chi}^+\tilde{l}$ loops.}
  \label{fig:fdchiloop}
\end{figure}

Still, one might
wonder if the branching ratio into the photon could
be modified via loops containing SUSY particles\footnote{For the
calculation we have used the formulas given in  \cite{Haber:1988px} for the
$H^\pm$-$\tilde \chi^\mp_i$ loops for 
$\tilde \chi^0_i \to \tilde \chi^0_j \gamma$
substituting correspondingly masses and couplings.}, 
such as those shown in
\cref{fig:fdchiloop}. We have varied the soft SUSY parameters
in the ranges $450 \le m_{\tilde L} \le 1000$~GeV, 
$200 \le m_{\tilde E} \le 1000$~GeV, -1 $\le A_\tau \le 1$~TeV,
200~$\le M_2\le$ 1000~GeV, $5 \le \tan\beta\le 40$, as well as 
$100 \le |\mu| \le 500$~GeV. We find 
at most a 10 per-cent variation in the branching ratio
for scenarios with light sleptons and charginos,
and small $\tan\beta$. The main reason for this is due to the experimental
bounds on the masses of the SUSY particles, which imply that the corresponding loops are suppressed with respect to the $W$-$l$ contributions. 
Thus, we conclude that the presence of SUSY does not affect the NuStar bounds.

In light of this negative result, one has two options to avoid the bounds. 
The first option is to increase $M_4$ above 700 MeV, such that $\nu_4$ decays quickly, without affecting Big Bang Nucleosynthesis (BBN)~\cite{Vincent:2014rja}. The second option is to decrease $m_1$ 
under $10^{-9}\,$eV, which suppresses the $\nu_4$ Yukawa coupling (see 
\cref{eq:Yuks}). This in turn decreases the mixing with the active neutrinos. 
Nevertheless, in both options $\nu_4$ can no longer be a good dark matter 
candidate, in the first case it decays too soon, while on the second case it is unlikely to be produced\footnote{One might wonder if the presence of SUSY could lead to more efficient production processes, for example due to  $\tilde \nu_R$ decays~\cite{Shakya:2016oxf}. We have checked in~\cite{Cerna-Velazco:2017cmn} that in our model SUSY particles have negligible branching ratios into $\nu_R$ states, so one can safely disregard heavy neutrino production through SUSY decays.} via oscillations with
the left-handed neutrinos~\cite{Dodelson:1993je,Shi:1998km}.
We thus discard $\nu_4$ as a candidate for dark matter in this model.

\begin{table}[t]
{\centering
 \begin{tabular}{c | c }
$M_4$ (keV) &  $\sin^2  2\theta$\\
\hline
7            &  $1.4\times10^{-7}$\\
30             & $3.3\times10^{-8}$\\
50             &$2.0\times10^{-8}$\\
\end{tabular}}
\caption{Predicted $\sin^2 2\theta$, for different values of $M_4$.}
\label{tab:mixcompare}
\end{table}

\subsection{Sneutrino Dark matter}

The calculation of the $\tilde\nu_1$ relic density depends on whether it is in thermal equilibrium with the primordeal plasma, or not. For this to happen, one requires that at some temperature $T$ we have:
\begin{equation}
\label{eq:condition}
H(T)<\langle\sigma v\rangle_T\, n(T)~,
\end{equation}
where $H(T)$ is the Hubble constant, $\langle\sigma v\rangle_T$ is the thermal average of the cross-section times velocity, and $n(T)$ is the number density of the $\tilde\nu_1$. Ultimately, a relevant factor in this condition is the size its couplings with other thermal particles, which are in turn determined by the size of the $Y_\nu$.

If the $\tilde\nu_1$ couplings are large enough such that it can be in thermal equilibrium at some temperature, the final value of $\Omega h^2$ follows the \textit{freeze-out} mechanism, see e.g.~\cite{Bertone:2004pz} and refs.~therein. Here, the expansion of the Universe leads to a point where \cref{eq:condition} does not hold, and the $\tilde\nu_1$ decouples from the plasma. The relic density depends on the number density at the freeze-out temperature, as well as on (co-)annihilations with other thermal particles.

If the couplings are too small, the $\tilde\nu_1$ relic density proceeds from decays or annihilation of other particles in the primordeal plasma. In the former case, the value of the relic density is mainly due to decays of the
next to lightest SUSY particle
(NLSP). For example, thermal neutralinos could decay after they freeze-out via $\tilde\chi^0\to\tilde\nu_1\,\nu$. In this situation the $\tilde\nu_1$ is said to follow the \textit{super-WIMP} mechanism~\cite{Feng:2003xh,Feng:2003uy,Asaka:2005cn,Asaka:2006fs}, and its relic density is proportional to the NLSP yield after freeze-out. In contrast, if the annihilations are more important than decays, the mechanism is that of \textit{freeze-in}~\cite{Hall:2009bx}. Here, the $\tilde\nu_1$ is created throughout the thermal history of the Universe by annihilations of all thermal particles, not only the NLSP. 
In both of the scenarios described above, the $\tilde\nu_1$ is called a feebly interacting massive particle (FIMP), 
and one assumes that it was not generated by other means at earlier times, such as during an inflationary period.

The naive seesaw expectation is that a light $\tilde\nu_1$ should have very small $Y_\nu$ couplings, and should thus be non-thermal. An important result of this procedure is that the NLSP can be long lived, which can lead to displaced vertices, disappearing tracks or heavy metastable charged particle signals at colliders~\cite{deGouvea:2006wd,Banerjee:2016uyt,Belanger:2018sti}.

If the neutrino mass generation mechanism allows for larger Yukawas, the $\tilde\nu_R$ can be thermal~\cite{Gopalakrishna:2006kr}. From 
\cref{eq:condition}, we see that as long as $\langle\sigma v\rangle_T\, n(T)/H(T)>1$ for some value of $T$, these particles will have been in thermal equilibrium at some point in the history of the Universe. Thus, as a first step, we evaluate the required size of $Y_\nu$ for this condition to hold.

To this end, we used \texttt{SARAH 4.14.0}~\cite{Staub:2008uz,Staub:2013tta,Staub:2012pb,Staub:2010jh,Staub:2009bi} to implement the model of~\cite{Cerna-Velazco:2017cmn} in \texttt{SPheno 4.0.3}~\cite{Porod:2003um,Porod:2011nf}, which calculates the mass spectrum and branching ratios. We used \texttt{SSP 1.2.5}~\cite{Staub:2011dp} to carry out the parameter variation. With the output, the variable $\langle\sigma v\rangle_T\, n(T)/H(T)$ was calculated with \texttt{Micromegas 5.0.9}~\cite{Belanger:2018mqt}. Here one has to keep two details in mind. First of all, the program only provides $\langle\sigma v\rangle^{\rm ann}_T$, the (co-)annihilation of $\tilde\nu_R$ with themselves and other SUSY particles into SM final states. In addition, \texttt{Micromegas} provides the total $n_{\rm eq}(T)$ from all SUSY particles, instead of the exclusive one for $\tilde\nu_1$. We do not consider this a serious problem, as $n_{\rm eq}(T)$ is dominated by the LSP. Still, in order to be certain of our results, in the following we shall consider $\langle\sigma v\rangle^{\rm ann}_T n_{\rm eq}(T)/H(T)<0.1$ as non-thermal, and $\langle\sigma v\rangle^{\rm ann}_T n_{\rm eq}(T)/H(T)>10$ as thermal.

\begin{figure}[t]
\includegraphics[width=0.495\textwidth]{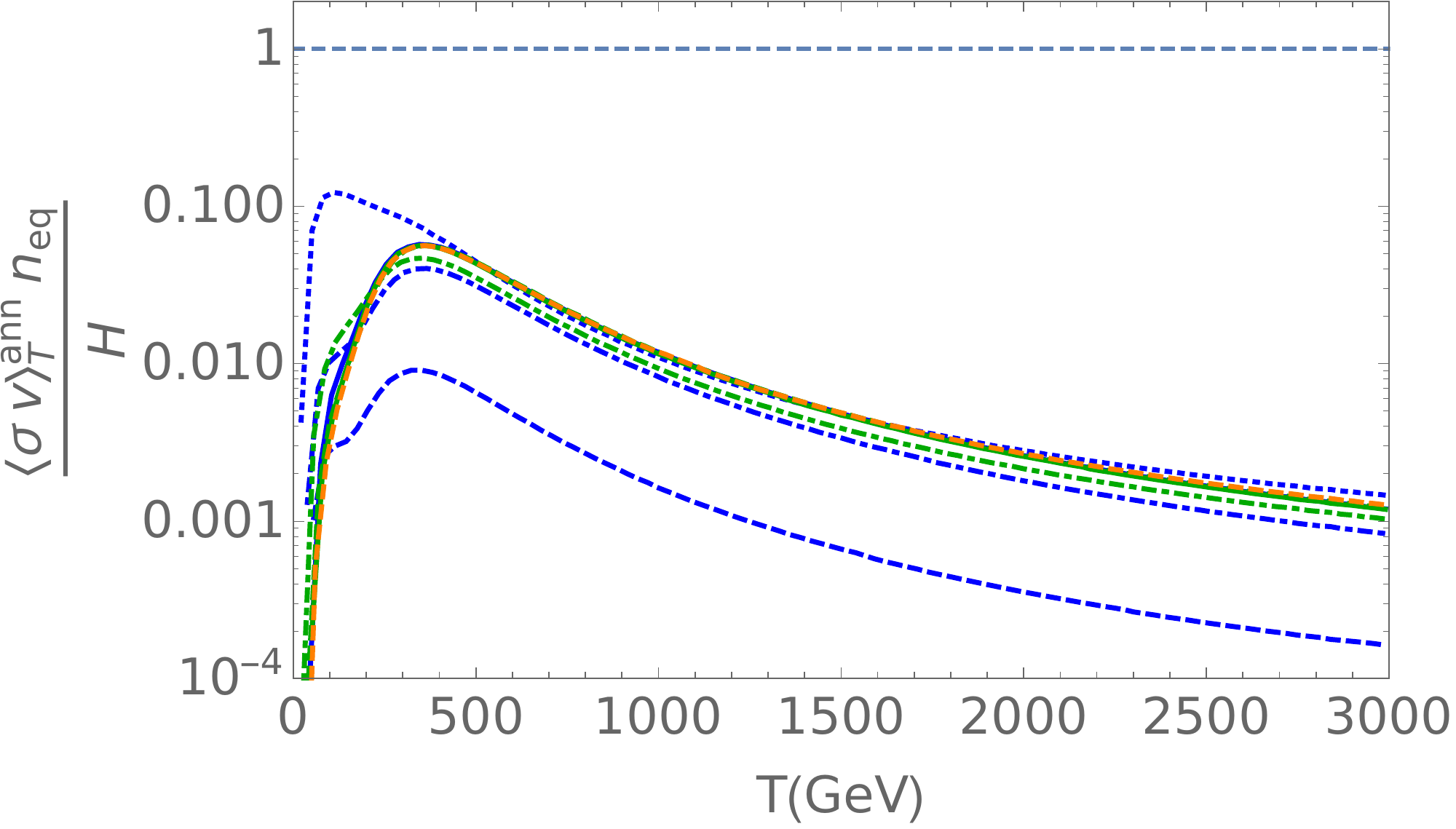}
\includegraphics[width=0.495\textwidth]{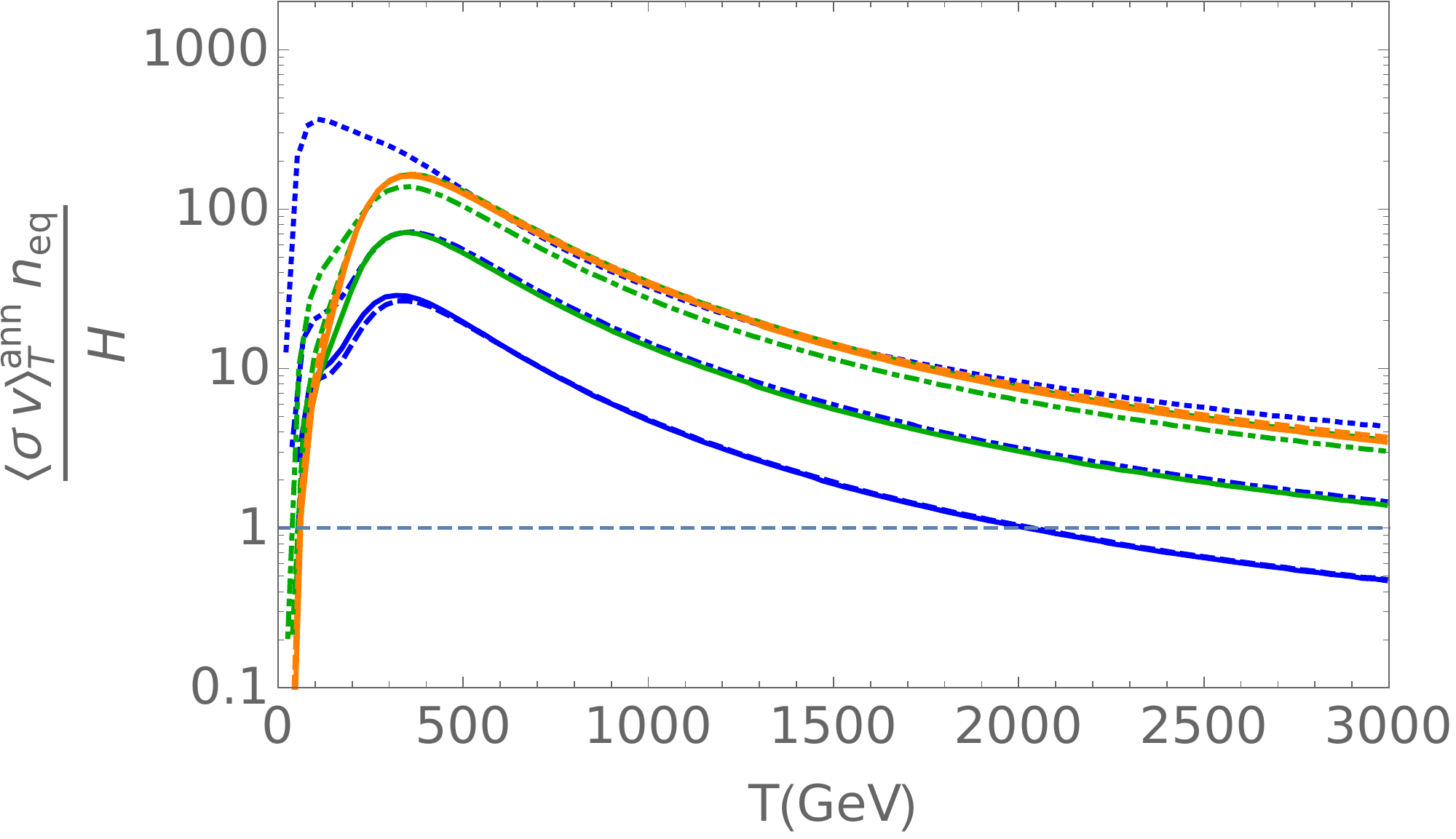}
\caption{Values of $\langle\sigma v\rangle_T^{\rm ann} n_{\rm eq}(T)/H(T)$ as a function of the temperature, for $Y_\nu^{\rm max}\sim\ord{10^{-7}}$ 
($\ord{10^{-5}}$) on the left (right). Blue, green and orange lines have 
$m_{\nu_R}=100,\,200,\,300$~GeV, respectively. L-slepton masses of 150, 
250, 350, 450~GeV are shown in dotted, dot-dashed, dashed and solid 
lines, if applicable.
}\label{fig:Temp}
\end{figure}
Our results are shown in \cref{fig:Temp}, where we evaluate $\langle\sigma v\rangle^{\rm ann}_T n_{\rm eq}(T)/H(T)$ as a function of $T$. Curves are shown for different SUSY masses, as well as different Yukawa couplings. We compare $Y_\nu^{\rm max}\sim\ord{10^{-7}}$ ($\gamma_{56}=3$) with $Y_\nu^{\rm max}\sim\ord{10^{-5}}$ ($\gamma_{56}=7$) on the left and right panels, respectively. We fix the SUSY masses as in \cref{Table:Parameters}, and probe $\tilde\nu_R$ soft masses of 100, 200 and 300 GeV (blue, green and orange, respectively). We also include MSSM slepton soft masses between 150 and 450~GeV, with the restriction $m_{\tilde L}=m_{\tilde E}>m_{\tilde\nu_R}$, such that the R-sneutrino is always the LSP.

In the Figure, we find that all curves of $\langle\sigma v\rangle^{\rm ann}_T n_{\rm eq}(T)/H(T)$ increase very quickly with $T$ up to a peak, with a subsequent drop as temperatures increase. This decrease is not surprising, since $H\sim T^2$, $n_{\rm eq}\sim T^3$ and $\langle\sigma v\rangle_T$ is expected to fall as $1/T^{2}$ for large temperatures. The conclusions of the Figure are straightforward. We find that, as long as $Y_\nu^{\rm max}\gtrsim\ord{10^{-5}}$, a $\tilde\nu_R$ can be thermal. This conclusion is valid for $\ord{100\,{\rm GeV}}$ scale R-sneutrino and L-slepton masses. For lower values of $Y_\nu$ a more careful analysis would be necessary, unless $Y_\nu^{\rm max}\lesssim\ord{10^{-7}}$, where we find no temperature which satisfies the condition $\langle\sigma v\rangle_T^{\rm ann} n_{\rm eq}(T)/H(T)>1$ and the $\tilde\nu_R$ is certainly a FIMP.

In the following, we explore the status of DM in the setup of~ref.~\cite{Cerna-Velazco:2017cmn}. As mentioned in the Introduction, we seek to enhance Yukawa couplings in order to avoid bounds on long-lived particles, and to hope for experimental correlations between the $\nu_R$ and $\tilde\nu_R$ sectors. To achieve this, we set $\gamma_{56}=7$, so $\tilde\nu_{R5}$ and $\tilde\nu_{R6}$ are \textit{thermal} particles. For simplicity, we also fix $m_{\tilde\nu_{R5}}=m_{\tilde\nu_{R6}}$. In contrast, the remaining eigenstate $\tilde\nu_{R4}$ is left with small Yukawa couplings, such that it is a \textit{non-thermal} $\tilde\nu_R$.

If the $\tilde\nu_{R4}$ mass is larger than that of $\tilde\nu_{R5}$ or $\tilde\nu_{R6}$, then the latter is a thermal LSP, and the relic density $(\Omega h^2)^{\rm th}$ is generated via freeze-out. In this case we find that the value of $(\Omega h^2)^{\rm th}$ calculated by \texttt{Micromegas} is extremely large, ruling out the whole parameter space.

Alternatively, if $\tilde\nu_{R4}$ is the LSP, the relic density is obtained from a combination of super-WIMP 
$(\Omega h^2)^{\rm dec}$ and freeze-in $(\Omega h^2)^{\rm in}$
processes\footnote{For the case of a generic fermionic DM candidate see
\cite{Garny:2018ali}.}. 
The super-WIMP contribution is obtained from\footnote{\texttt{Micromegas} assumes that the decay products of new non-DM particles thermalize instantly, so entropy-dilution effects such as those in~\cite{Asaka:2006ek} are not included.}:
\begin{equation}
(\Omega h^2)^{\rm dec}=(\Omega h^2)^{\rm th}\frac{m_{\tilde\nu_1}}{m^{\rm th}_{NLSP}}
\end{equation}
where $m^{\rm th}_{NLSP}$ is the mass of the lightest thermal SUSY partner. Thus, one can adjust this contribution to any desired value by choosing an appropriate $m_{\tilde\nu_1}$. In this case it is not necessary for the thermal $\tilde\nu_R$ to be NLSP, it is perfectly possible to have a thermal $\tilde\nu_L$ or $\tilde h$ NLSP which later decays to the $\tilde\nu_1$.

The super-WIMP contribution by itself cannot explain the observed relic density in two situations. First, since $(\Omega h^2)^{\rm dec}\leq(\Omega h^2)^{\rm th}$, it is impossible to get the correct relic density if $(\Omega h^2)^{\rm th}$ is too small (this can be the case for both $\tilde\nu_L$ and $\tilde h$ NLSP). Second, it is also possible for $(\Omega h^2)^{\rm th}$ to be so large that the feeble $\tilde\nu_1$ mass must be lower than the $\nu_4$ mass. This situation would require a negative soft $m^2_{\nu_{R4}}$, which we shall discard\footnote{Strictly
speaking, this only true in case of diagonal $m_{\tilde\nu_R}^2$. However, to decrease the $\tilde\nu_{R4}$ mass below $M_4$ would require very large off-diagonals, with the risk of giving large couplings to all $\tilde\nu_R$ 
and/or substantial contributions to lepton flavour violating
observables, which are severely bound by experimental data.}. 
Nevertheless, only the latter case is a problem, since a small $(\Omega h^2)^{\rm dec}$ can be complemented by a larger $(\Omega h^2)^{\rm in}$.

\begin{figure}[tb]
\includegraphics[width=0.55\textwidth]{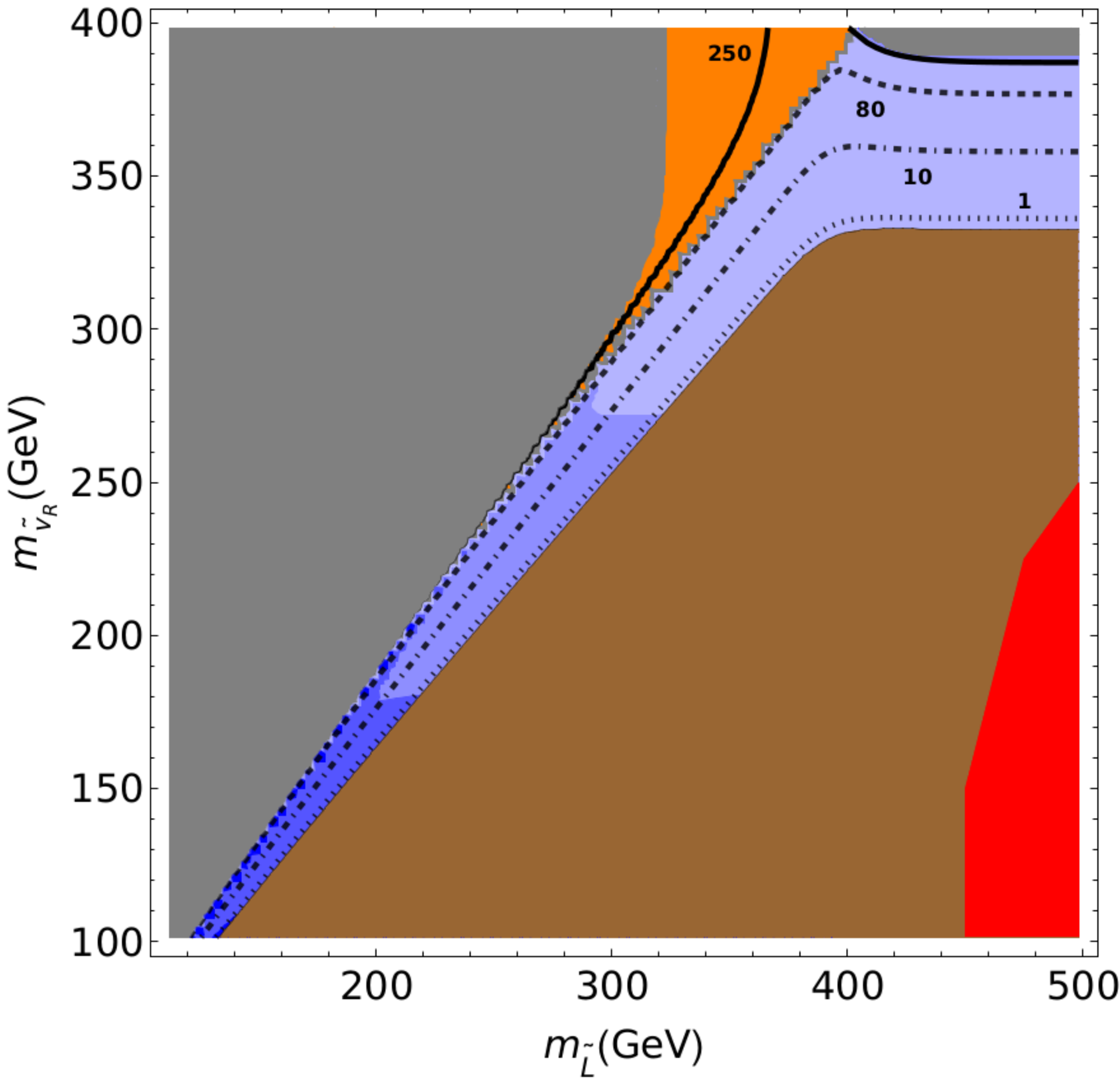}
\caption{\label{fig:FIMPmass} Maximum possible mass of feebly interacting sneutrino, such that $(\Omega h^2)^{\rm dec}$ matches the observed relic density. Masses of 250, 80, 10 and 1~GeV are shown in solid, dashed, dashed-dotted and dotted black lines. Lifetimes smaller than $10^4$, $10^5$ and $10^6$ seconds are shown in light blue, blue and dark blue, respectively. Brown and red regions are excluded, gray regions cannot reproduce the correct relic density using Super-WIMP mechanism. More details are given in the text.} 
\end{figure}
We show the results of this super-WIMP-only analysis in \cref{fig:FIMPmass}.
We vary $m_{\tilde L}=m_{\tilde E}$ and $m_{\tilde\nu_{R}}\equiv m_{\tilde\nu_{R5,6}}$, which are the soft masses of thermal particles, and assume $m_1=10^{-3}$~eV and $M_4=700$~MeV. The black contour lines shown the 
maximum allowed mass of $\tilde\nu_1$, such that the super-WIMP mechanism 
saturates the relic density, that is, $(\Omega h^2)^{\rm dec}=0.12$. We 
find that $m_{\tilde\nu_1}$ spans from 700~MeV to almost 300~GeV.

On the gray regions of the Figure, the thermal L-sleptons and higgsinos have large \mbox{(co-)annihilation} cross-sections, leading to a too small $(\Omega h^2)^{\rm th}$. This means that the super-WIMP mechanism cannot explain by itself the observed relic density, so there is no upper bound on $m_{\tilde\nu_1}$ apart from the requirement of being the LSP.

The brown and red regions are excluded. On the brown region, the feeble
$\tilde\nu_1$ requires a mass lower than the minimum of $M_4=700$~MeV, so $(\Omega h^2)^{\rm dec}$ is too big.  Moreover, the red region was ruled out by LHC searches in Ref.~\cite{Cerna-Velazco:2017cmn}. With this, we conclude that if the Super-WIMP mechanism was entirely responsible for dark matter production, then it would only be viable in a reduced region of the evaluated parameter space.

The main annihilation channels giving $(\Omega h^2)^{\rm th}$ vary throughout the Figure. For $m_{\tilde L}$ below $\mu=400$~GeV, $\tilde\nu\tilde\nu$ annihilation dominates, mainly into $W^+W^-$, $Z\,Z$ and $t\,\bar t$ final states. Above 400 GeV, neutralino and chargino (co-)annihilation determine the thermal relic density, with channels into $u_i\,d_i$ and $e_i\,\nu_j$ final states giving the largest contributions.

Due to $\tilde\nu_1$ being a FIMP, it is important to be aware of the thermal NLSP lifetime. If longer than one second, it can be subject to BBN
constraints~\cite{Kawasaki:2004qu,Ishiwata:2009gs,Banerjee:2016uyt,Kawasaki:2017bqm}. This information is included in \cref{fig:FIMPmass}.
On the Figure, the blue regions feature $\tilde\nu_{R5,6}$ NLSP lifetimes between $10^3$ and $10^6$ seconds.  On the orange region, the $\tilde\nu_L$ is the NLSP, with a lifetime smaller than one second. In the gray regions we find that, for a 700 MeV $\tilde\nu_1$, the lifetime of the $\tilde\nu_L$ NLSP (left region) is also lower than a second. However, on the upper right gray region we again have a $\tilde\nu_R$ NLSP, with a lifetime of around $10^2$~s.

It is important to take into account that this information on the lifetime is strictly valid for the selected values of $m_1$ and $M_4$, since the $|(Y_\nu)_{a4}|^2$ coupling of $\tilde\nu_{R4}$ depend on them (see \cref{eq:Yuks}). For smaller $m_1$ or $M_4$, it will increase. We briefly discuss the impact of the NLSP lifetime on BBN further below.

We now turn to the calculation of the freeze-in contribution, $(\Omega h^2)^{\rm fi}$, which is again calculated with \texttt{Micromegas}. Throughout the evaluated parameter space, we find that if we take $m_{\tilde\nu_1}$ equal to the maximum value allowed by the super-WIMP mechanism, the freeze-in contribution can greatly exceed the observed value.

\begin{figure}[tb]
\includegraphics[width=0.65\textwidth]{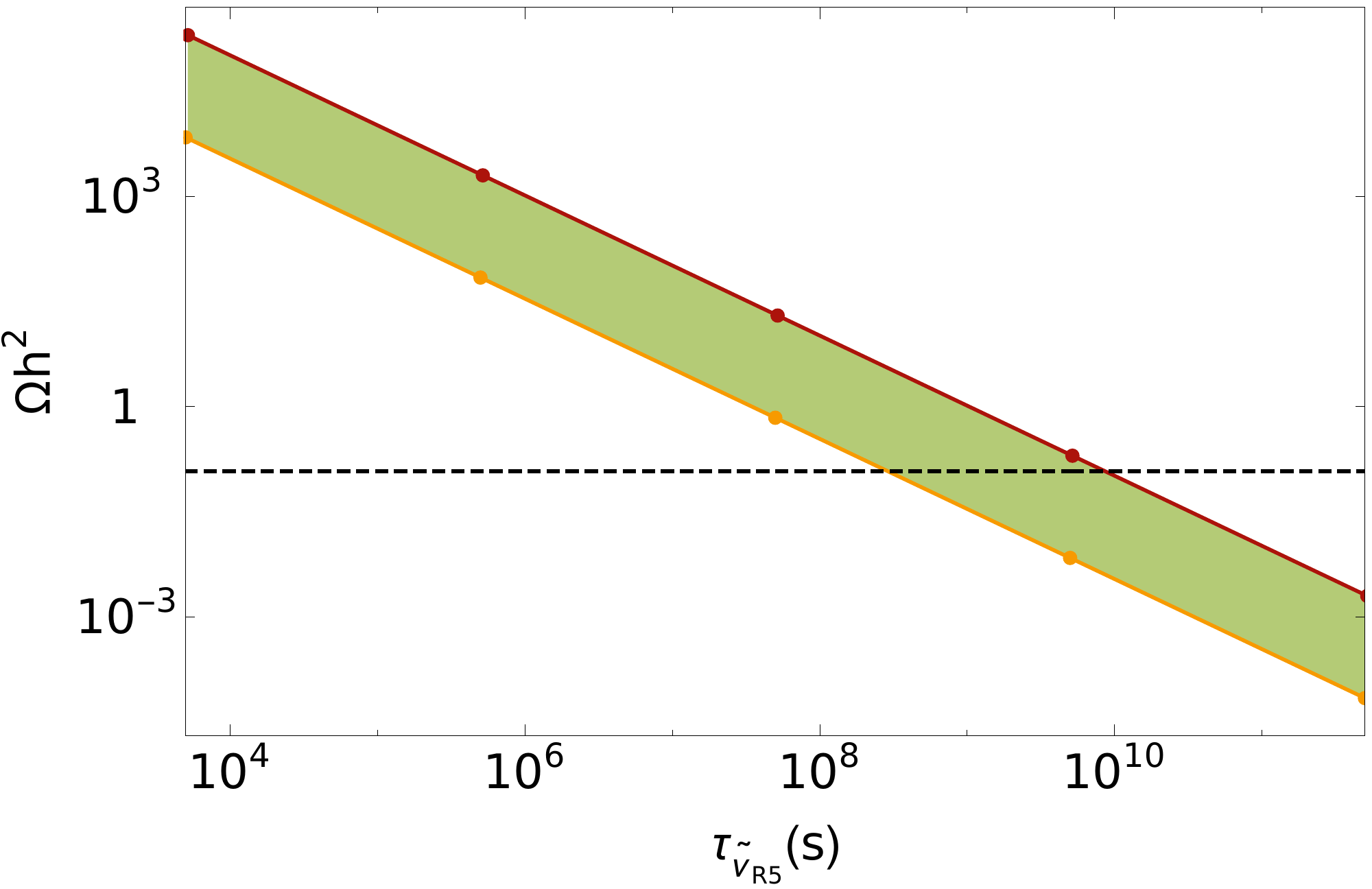}
\caption{\label{fig:freezein} Freeze-in contribution to our test point. Red (yellow) curve shows contributions for maximum (minimum) $\tilde\nu_1$ mass. Dashed line gives observed $\Omega h^2$. The dots indicate $m_1=10^{-3},\,10^{-5},\,10^{-7},\,10^{-9}$ and $10^{-11}$~eV from left to right.} 
\end{figure}

One way of suppressing freeze-in is by reducing the lightest neutrino mass $m_1$, which decreases $(Y_\nu)_{a4}$. Due to the long calculation time, we show results for only one representative point in \cref{fig:freezein}, with $M_4=700$~MeV, $m_{\tilde L}=323$~GeV and $m_{\tilde\nu_R}=302$~GeV, where $\tilde\nu_{R5}$ is the NLSP. We see that reducing $(Y_\nu)_{a4}$ increases the $\tilde\nu_{R5}$ lifetime, as shown in the red curve of the Figure. Given the large lifetime, assuming that BBN constraints are avoided, it is still important to check that the time for recombination $\sim10^{13}\,$s is not exceeded, else the $\tilde\nu_{R5}$ would decay into $\tilde\nu_1$ after the cosmic microwave background was emitted. In this case, the relic density observed by Planck~\cite{Aghanim:2018eyx} would correspond to $(\Omega h^2)^{\rm th}$, ruling out the scenario. For our example, we find that $(\Omega h^2)^{\rm fi}$ becomes subdominant when $m_1\lesssim10^{-9}$~eV, and that the recombination constraint is satisfied. 
This solution is also attractive as no additional $\nu_R$ DM is produced, since both $M_4$ is large and $(Y_\nu)_{a4}$ is tiny.

Alternatively, it is possible to have a smaller $m_{\tilde\nu_1}$, such that the super-WIMP contribution is negligible. In this case, one needs to adjust the lightest neutrino mass, such that $(\Omega h^2)^{\rm fi}$ reproduces the observations. Such a scenario is shown on the yellow curve of \cref{fig:freezein}, where we have chosen $m_{\tilde\nu_1}=M_4=700$~MeV. Here one obtains the correct relic density when $m_1\approx\ord{10^{-8}}$~eV.

Thus, we find that in in this scenario, where $(Y_\nu)_{a4}$ is suppressed and $(Y_\nu)_{a5}$, $(Y_\nu)_{a6}$ are enhanced, the lightest R-sneutrino can reproduce the observed relic density, with a dominant super-WIMP or freeze-in mechanism depending on $m_{\tilde\nu_1}$ and $(Y_\nu)_{a4}$. However, regardless of the dominating mechanism, the NLSP lifetime is considerably large, and needs to pass BBN constraints. As the $\tilde\nu_{R5}$ has no hadronic decays, this model does not lead to any any hadro-dissociation nor $p\leftrightarrow n$ conversion processes. However, around $50\%$ of the decays are into charged leptons, so it could be subject to photo-dissociation processes. According to~\cite{Kawasaki:2017bqm}, these are important for lifetimes above $10^4-10^6$ seconds. Even though a solid conclusion can only be obtained through a detailed numerical simulation, which is outside the scope of this work, we can conclude that BBN constraints suggest that the $\tilde\nu_{R5}$ density, and thus $(\Omega h^2)^{\rm th}$, would need to be heavily suppressed. If confirmed, the only way to have R-sneutrino dark matter would be to either have a $\tilde\nu_L$ NLSP with a short lifetime (orange and upper left gray regions of \cref{fig:FIMPmass}) or to restrict ourselves to the upper right gray region of the Figure where $(\Omega h^2)^{\rm th}$ is small. Alternatively, we could decrease $(Y_\nu)_{a5}$ and $(Y_\nu)_{a6}$ such that no $\tilde \nu_R$ is thermal.

As a final remark, we must comment that other works attempted to have a thermal $\tilde\nu_R$ by generating a large mixing with the $\tilde\nu_L$~\cite{Arina:2007tm,Belanger:2010cd,Dumont:2012ee,Chatterjee:2017nyx}. The main idea was to allow the mass eigenstate to interact via gauge interactions, with a relative suppression of the overall coupling due to the mixing. This would allow the sneutrino to remain thermal, but at the same time avoid direct detection bounds. For this approach to work, it was required to have $\sin\tilde\theta\gtrsim10^{-2}$.

As we see from \cref{eq:LRmixing}, the L-R mixing is suppressed by 
$Y_\nu$. It is clear that even if we increased $\gamma_{ij}$ to their 
maximum values allowed by experiment, an additional enhancement is 
required to bring the mixing to the required level. Earlier works have 
considered an extremely large $A_\nu$, such that $Y_\nu A_\nu$ becomes of 
the order of the other soft terms~\cite{ArkaniHamed:2000bq}, e.g.~ the 
slepton mass parameters. 
However, this creates the danger of a charge breaking minimum similar
to the well-known problem of 
charge and color breaking minima within the MSSM, see e.g.~\cite{Casas:1995pd} for an extensive discussion on
tree-level constraints. 

In our case, we need to assess if the addition of the new $\hat\nu_R$ can lead scalar fields other than the Higgs to acquire a vacuum expectation value (\vev). In particular, we will evaluate the following \vev pattern for one generation of sleptons and sneutrinos:
\begin{align}
\langle H^+ \rangle = \langle \tilde l_L^-\rangle= 
\langle \tilde \nu_R \rangle =\alpha~,
\end{align}
as this corresponds to a $D$-flat direction in the scalar potential.
Assuming that all other fields do not acquire a \vev and that all
parameters are real,
one gets for the potential at tree level
\begin{align}
V = \left(m^2_{H_u} + \mu^2 + m^2_{\tilde L} + m^2_{\tilde \nu_R} 
+ M_R^2+ B_{\tilde \nu} \right) \alpha^2 
- 2 (T_\nu  + M_R^* Y_\nu) \alpha^3 + 3Y_\nu^2 \alpha^4
\end{align}
In order to avoid this direction in the potential from being equal or 
lower than zero, a sufficient way is  requiring a negative discriminant
in the solutions for $\alpha$ of the equation $V(\alpha)=0$. 
We get the following
bound\footnote{This result coincides with the one in ref.~\cite{Kakizaki:2015nua} in the limit of vanishing $M_R$ and  $B_\nu$
where the case of Dirac neutrinos has been studied. Note that their
$A_\nu$ coincides with our $T_\nu$.}
\begin{align}
(T_\nu + M_R Y_\nu)^2 \le 3 (m^2_{H_u} + |\mu|^2 + m^2_{\tilde L} + m^2_{\tilde \nu_R}  + M^2_R + B_{\tilde \nu}) Y_\nu^2
\end{align}
Writing $T_\nu = A_\nu Y_\nu$ we see immediately that 
\begin{align}
(A_\nu + M_R)^2 \le 3 (m^2_{H_u} + |\mu|^2 + m^2_{\tilde L} + m^2_{\tilde \nu_R}  + M^2_R + B_{\tilde \nu}) 
\end{align}
which implies it is $A_\nu$, and not $T_\nu$, who can be at most of the order of soft terms. Thus L-R mixing is not a safe procedure to solve the DM problem in this
context. We note for completeness that this tree-level estimate
can get significant loop corrections, as has been in shown
in similar models~\cite{Camargo-Molina:2013sta,CamargoMolina:2012hv}. 
However, the estimate gives
the correct order of magnitude, which is sufficient in our case.
We have explicitly checked that this estimate is correct
within a factor two for a few points using the package \texttt{Vevacious}~\cite{Camargo-Molina:2013qva}.

\section{A $U(1)_{B-L}$ extension}
\label{sec:mod_BL}

We have seen in the previous Section that in the minimal
model it is not possible to have a thermal $\tilde\nu_R$ as an LSP, as it will yield a too large contribution to $\Omega h^2$.
We therefore seek an extension where the R-sneutrino can be a thermal relic, potentially giving the correct
relic density, without impacting too much the high-energy and precision phenomenology,
 in particular in view of the collider constraints obtained in
 \cite{Cerna-Velazco:2017cmn}. Several possibilities have
 been considered in a similar context in the literature: an additional
 $U(1)$ gauge factor
 \cite{Lee:2007mt,Belanger:2011rs,Bandyopadhyay:2011qm,Basso:2012gz},
left-right symmetric models \cite{Frank:2017tsm}, the  NMSSM
\cite{Cerdeno:2009dv,Cerdeno:2011qv} or via additional 
F- and D-terms as 
they occur in models of hybrid inflation \cite{Deppisch:2008bp}.

Taking a $U(1)_{B-L}$ extension as an example,
we will show how this can be achieved in the parameter space
we are interested in here. For this,
we first briefly summarize the main aspects 
of the model presented in~\cite{OLeary:2011vlq}. 
Here, the MSSM particle content is extended by three new types of superfields. First, one has a $B'$ vector superfield associated to the $U(1)_{B-L}$ symmetry. Second, one adds two new Higgs-like chiral superfields, $\hat\eta$ and $\hat{\bar\eta}$, 
carrying $B-L$ number $\pm2$, whose scalar components break $U(1)_{B-L}$ and provide mass to the $Z'$ boson. These chiral superfields will be called \textit{bileptons} in the following. Note, that the symmetry breaking is such that $R$-parity is conserved\footnote{This symmetry-breaking pattern is only a necessary but not a sufficient condition, as there could still be spontaneous $R$-parity violation due to a sneutrino \vev~\cite{CamargoMolina:2012hv}.}. Finally, the anomaly-cancellation 
requires the existence of three right-handed neutrino superfields $\hat\nu_R$~\cite{Barger:2008wn}. These have masses around $10$-$100$~GeV if the $U(1)_{B-L}$ breaking scale is of the order 1-100~TeV.

The superpotential is given by 
\begin{equation}
\mathcal{W}= \mathcal{W}_{\rm MSSM} 
  - \mu' \,\widehat{\eta}\,\widehat{\bar{\eta}}\,
+(Y_x)_{ij}\,\widehat{\nu}_{Ri}\,\widehat{\nu}_{Rj}\,\widehat{\eta}\,
+(Y_{\nu})_{ij}\,\widehat{L}_i\cdot\widehat{H}_u\,\widehat{\nu}_{Rj}
\label{eq:superpotBL}
\end{equation}
and one has the additional soft SUSY-breaking terms:
\begin{align}
\nonumber {\cal V}^{soft} = & {\cal V}_{\rm MSSM}^{soft}
 - {M}_{B B'}\lambda_{\tilde{B}} \lambda_{\tilde{B}'} 
 - \frac{1}{2}{M}_{B'}\lambda_{\tilde{B}'} \lambda_{\tilde{B}'}  
 - m_{\eta}^2\, |\eta|^2 - m_{\bar{\eta}}^2\, |\bar{\eta}|^2
 - B_{\mu'}\,\eta\, \bar{\eta}  \nonumber \\
& - ({m_{\tilde\nu_R}^2})_{ij} \tilde{\nu}_{Ri}^* \tilde{\nu}_{Rj}
+\bigg( 
 (T_{x})_{ij} \tilde{\nu}_{Ri}\,\tilde{\nu}_{Rj}\eta
 +(T_{\nu})_{ij} \tilde{L}_i\cdot H_u\,\tilde{\nu}_{Rj}
 +\text{h.c.}\bigg)~.
 \label{eq:LsoftBL}
\end{align}
 Without loss of generality 
one can take $B_\mu$ and $B_{\mu'}$ to be real, as
in the MSSM. The extended gauge group breaks to
$SU(3)_C \otimes U(1)_{em}$ as beside the Higgs fields also the
bileptons receive \vevs, denoted by $v_\eta$ and $v_{\bar{\eta}}$,
see ref.~\cite{OLeary:2011vlq} for details.
We define $\tan\beta' = \frac{v_{\eta}}{v_{\bar{\eta}}}$ in analogy to
the ratio of the MSSM \vevs.
In the following we will neglect all effects concerning gauge kinetic mixing as they do not impact our findings, e.g.\ we will set  $M_{BB'}=0$. 

As in the MSSM one can always get a SM-like light Higgs boson
$h$, with mass of 125 GeV, assuming third generation squarks sufficiently heavy and having a sizable mixing in the stop sector. Beside $h$ this model
contains a second scalar $h'$, which is light provided $\tan\beta'$ is close to 1~\cite{OLeary:2011vlq}. As the limits on $Z'$ are in the multi-TeV
range, this  state is mainly an SM gauge-singlet, since
the mixing scales like $v_u/\sqrt{v^2_{\bar{\eta}}+v^2_\eta}$. However, 
since $h$ can still have a non-negligible mixing with $h'$, we use \texttt{HiggsBounds} \cite{Bechtle:2013wla} to check if its properties are compatible with the experimental results.

Neglecting mixing effects between the MSSM Higgs bosons and the bileptons, one gets for the mass of $h'$ at tree level~\cite{OLeary:2011vlq}:
\begin{align}
m^2_{h'}= \frac{1}{2} \left( m^2_{Z'} + m^2_{A^0{}'}
- \sqrt{(m^2_{Z'} + m^2_{A^0{}'})^2- 4 m^2_{Z'} m^2_{A^0{}'} \cos^2 2 \beta' }\right)~,
\end{align}
which has a structure similar to the MSSM case\footnote{We have checked that loop corrections are less important in this sector compared to the MSSM part.}.  From the above
equation it follows that, for fixed $m_{Z'}$ and $m_{A^0{}'}$, one can control the value of $m_{h'}$ by adjusting $\tan\beta'$. In particular, $m_{h'}\to0$ when $\tan\beta'\to1$, and $m_{h'}\to{\rm Min}(m_{Z'},\,m_{A^0{}'})$ for  $\tan\beta' \to 0$ or $\tan\beta' \to \infty$.

Regarding the sneutrino sector, the $\Delta L=2$ operators, such as $(Y_x)_{ij}\,\hat{\nu}_{Ri}\,\hat{\nu}_{Rj}\,\hat{\eta}$, set, on the
one hand, the mass scale of the right-handed neutrinos and 
lead, on the other hand, to a splitting of the complex sneutrino fields into their CP-even and CP-odd components, denoted $\tilde \nu^S$ and 
$\tilde \nu^P$, respectively.
Neglecting left-right mixing and assuming CP conservation, the masses of the \rsnus can be expressed as \cite{Basso:2012gz}
\begin{align}
\label{eq:mSnuA}
\mReSnuSq \simeq & \,\, m_{\tilde\nu_R}^2
 + m^2_{Z'} \left( \frac{1}{4} \cos(2 \beta')
                 + \frac{2 Y_x^2}{\gBL{2}} \sin^2 \beta'\right) + m_{Z'} \frac{\sqrt{2} Y_x}{\gBL{}}
    \left(A_x \sin\beta'-\mu' \cos\beta' \right)\, ,\\
\mImSnuSq \simeq & \,\, m_{\tilde\nu_R}^2
 +  m^2_{Z'} \left( \frac{1}{4} \cos(2 \beta')
                 + \frac{2 Y_x^2}{\gBL{2}} \sin^2\beta' \right)
                 - m_{Z'}  \frac{\sqrt{2} Y_x}{\gBL{}}
    \left(A_x  \sin\beta'-\mu' \cos\beta' \right)\, .
\label{eq:mSnuB}
\end{align}

Dark matter aspects of this model have been discussed
in  \cite{Basso:2012gz} where various candidates
were evaluated, including sneutrinos. It has
been shown that in case of a CP-even sneutrino
it is possible to reproduce the correct relic density through a Higgs funnel mechanism, similar to
other related models (see discussion in~\cite{DeRomeri:2012qd}). This mechanism is also applicable to CP-odd sneutrinos, being a consequence
of the $D$-term and $F$-term couplings of  the $h'$ to
the $\tilde\nu_R$. 
One issue that we want to address here is whether the Higgs funnel mechanism still explains the observed $\Omega h^2$ if we enforce a small or even 
negligible mass splitting between the scalar and pseudoscalar sneutrino 
states.

As it is well know, funnel regions imply particular
relation between various masses, e.g.\ here we shall require $m_{\tilde{\nu}_1^S}\simeq m_{h'}/2$. Moreover, having
a small splitting between the scalar and pseudoscalar implies
$A_x \sin\beta' \simeq \mu' \cos\beta'$ as can be seen
from \cref{eq:mSnuA,eq:mSnuB}. We take
as input the mass of the pseudoscalar sneutrino 
$m_{\tilde \nu_1^P}$, $\tan\beta'$, $m_Z'$ and fix the
mass difference between $\tilde{\nu}_1^S$ and $\tilde{\nu}_1^P$
to 0.5 GeV. The values of the parameters, which we 
fix for these investigations, are given in \cref{tab:parametersBL}.

\begin{table}[t]
 { \centering
  \begin{tabular}{c c| c c| c c | c c}
    \hline
    parameter & value & parameter & value
    & parameter & value & parameter & value \\
    \hline
    $\mu'$ & 500 GeV &  $M_1$ & 724 GeV &  
    $Y_{\nu,11}$ &$10^{-8}$ & $Y_{x,11}$ & $10^{-5}$ \\ 
    $M_Z'$ & 4.20 TeV &  $M_2$ & 1.19 TeV & 
    $\tan\beta$ & 20 & $Y_{x,22}$ & $1.9 \cdot 10^{-2}$ \\ 
    $m_{A^0}$ & 6.20 TeV &   $M_3$ & 3.10 TeV & 
     $g_{B-L} $ & 0.55 &  $Y_{x,33}$ & $1.9 \cdot 10^{-2}$ 
  \end{tabular}
  }
  \caption{The values of the fixed parameters used in the  numerical examples of BLSSM,}
  \label{tab:parametersBL}
\end{table}

\begin{figure}
  \centering
  \includegraphics[scale=0.5]{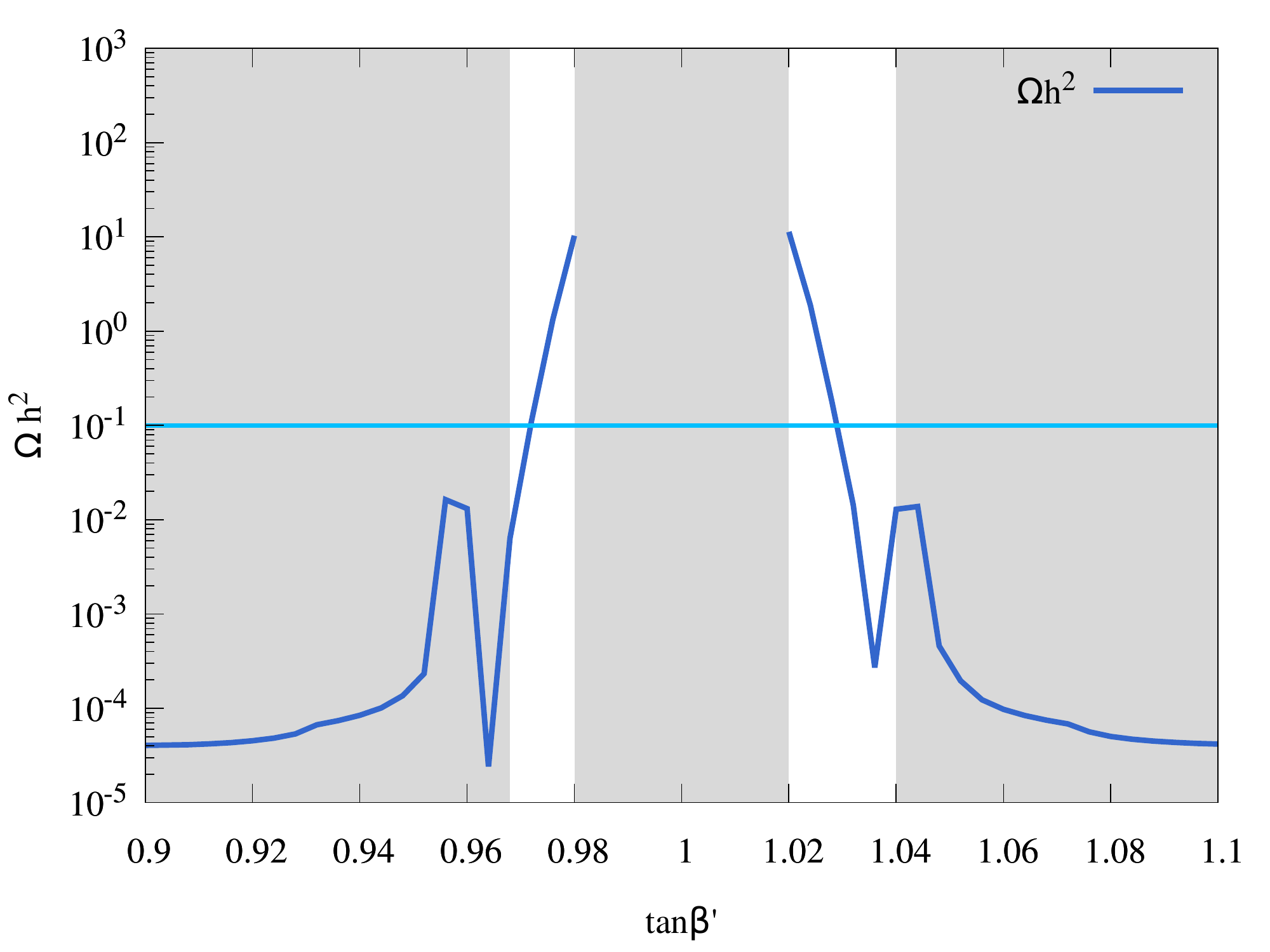}
  \caption{Relic density $\Omega h^2$ due to sneutrino dark matter as a function of
 $\tan\beta'$ for $m_{h'}= 2 m_{\tilde{\nu}^S_1}$ and
 $m_{\tilde{\nu}^S_1}-m_{\tilde{\nu}^P_1}=0.5$~GeV. 
 The other parameters are
 given in \cref{tab:parametersBL}. The two outer shaded areas are
 excluded by Higgs data whereas the middle one contains tachyonic
 states.}
  \label{fig:varytanb}
\end{figure}
As an example we show $\Omega  h^2$ as a function of $\tan\beta'$  in 
\cref{fig:varytanb}. In the Figure, at the boundaries of the evaluated 
values of $\tan\beta'$, we find $m_{h'}\approx350$~GeV and a  mixing of a 
few per-cent with the SM-like Higgs. Due to the mixing, several final 
states are allowed when the $\tilde\nu$s annihilate through the funnel. For larger masses of $h'$, the dominant annihilation is into W$^\pm$ pairs . 
As $\tan\beta'$ approaches unity, the co-annihilations are less efficient, as this channel gets suppressed and the $b\bar{b}$ final state becomes
the most important one. Close to the central region, excluded due to tachyonic states, this channel is also closed and the dominant process is
 $\tilde\nu\tilde\nu\to\nu_4\nu_4$. Since it is suppressed by a small 
 $Y_x$, the relic density increases considerably.

Thus, similar to the case of the MSSM, the Higgs funnel is
only possible within a small strip in parameter space. In 
the case shown we find exactly two working values of 
$\tan\beta'$. Interestingly enough, both are consistent with 
the experimental constraints, which we have checked using 
\texttt{HiggsBounds}. We note for completeness, that we find the soft 
parameter $m^2_{\tilde\nu_R}<0$ when $\tan\beta'<1$. However, when
matching this model to the one of the previous Section, the D-term
contribution from the $Z'$ yields a positive value.\footnote{We have checked
that also in the region with $m^2_{\tilde\nu_R}<0$ the R-parity
conserving minimum is the global one using \texttt{Vevacious }
\cite{Camargo-Molina:2013qva}.}  In the points explaining the relic 
density, $h'$ has about one per-cent admixture with the SM-like 
Higgs boson.

\begin{figure}[t]
  \centering
    \includegraphics[scale=0.32]{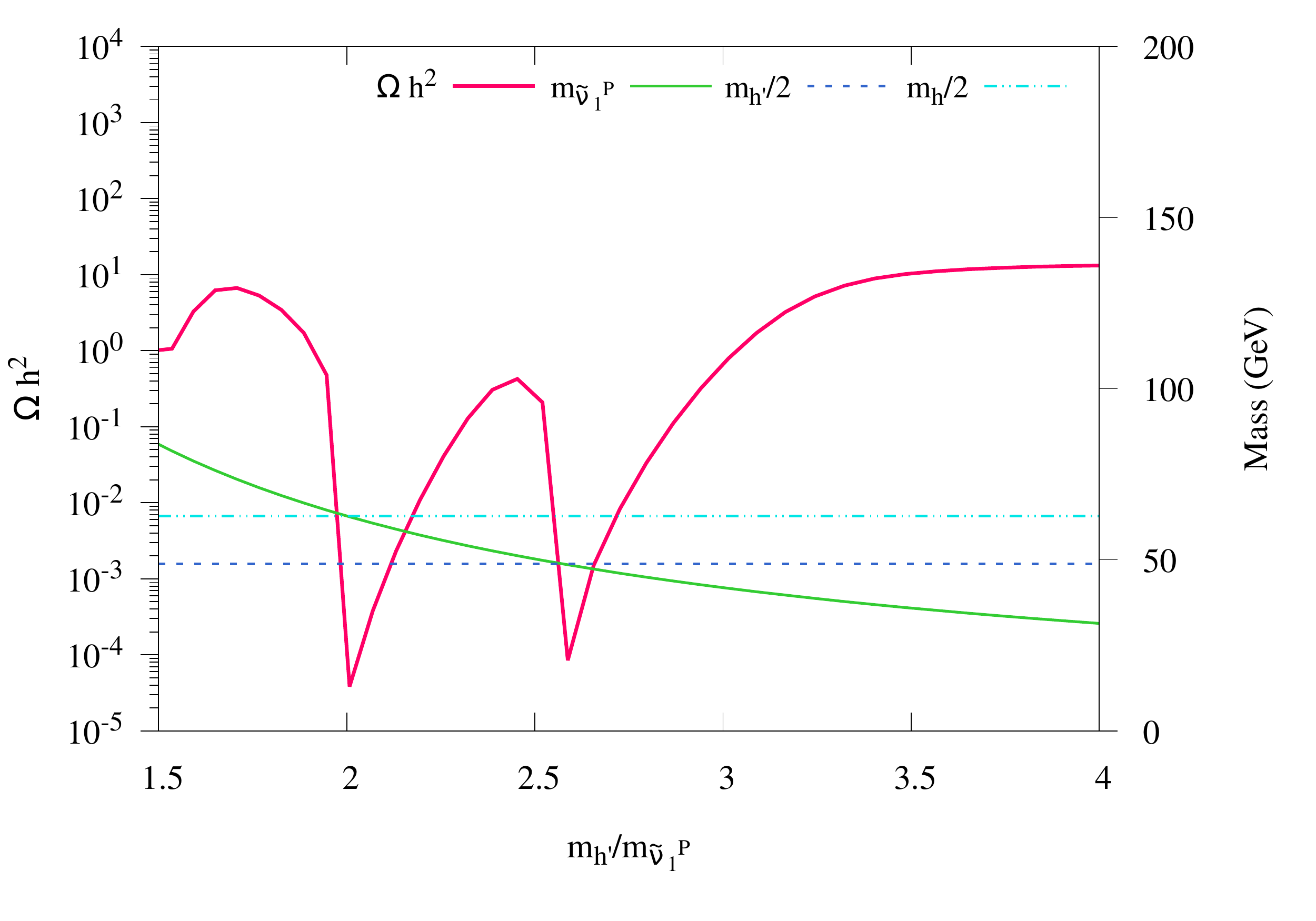}
    \includegraphics[scale=0.32]{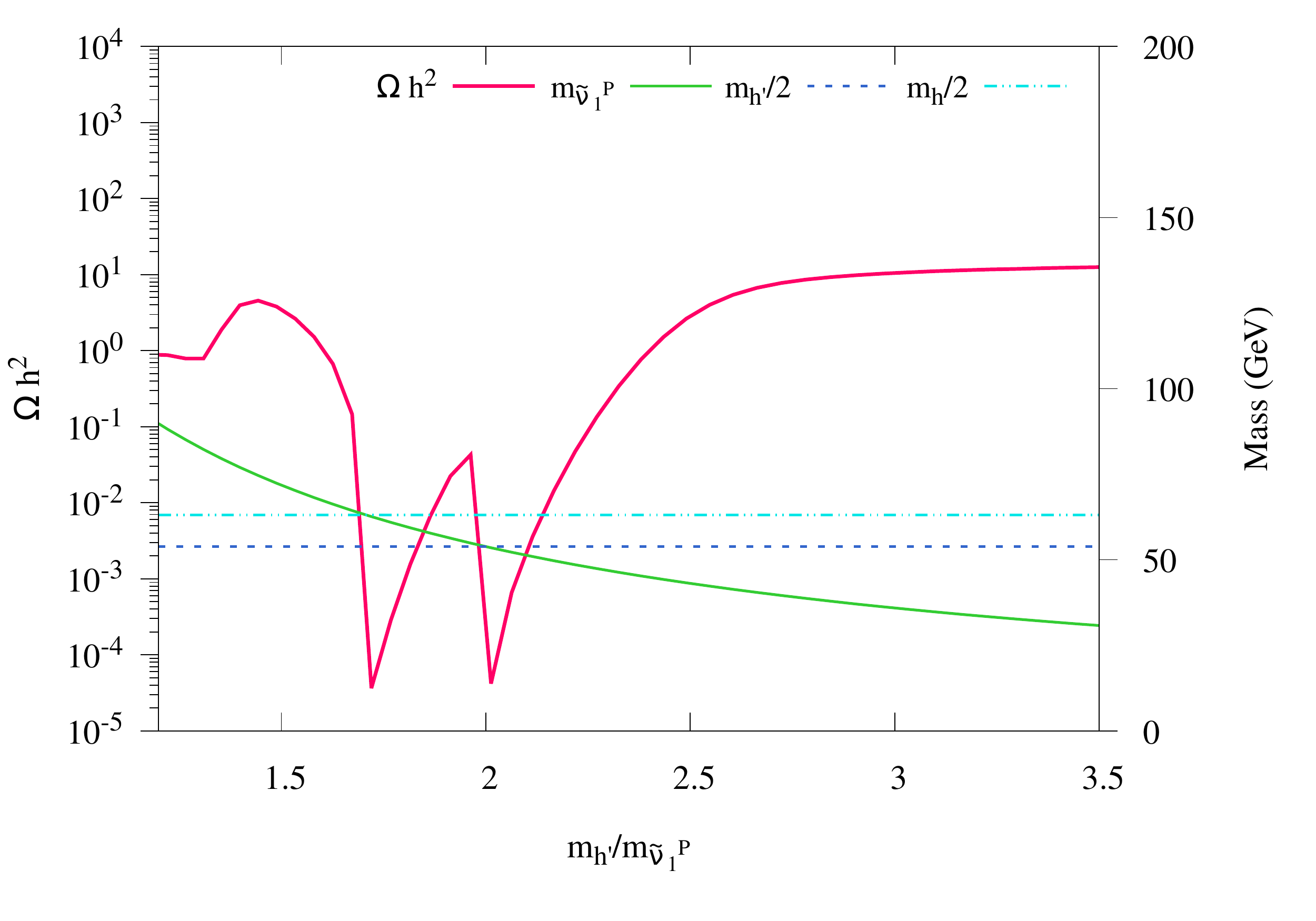}
    \caption{Relic density $\Omega h^2$ due to sneutrino dark matter as a function 
of  mass ratio $m_{h'}/m_{\tilde{\nu}_1^P}$. We have set
$m_{\tilde{\nu}_1^S}=m_{\tilde{\nu}_1^P}+0.5$~ GeV,
$\tan\beta'= 0.972$ (left plot) and $\tan\beta'= 1.032$ (right plot).
On the left (right) plot we find $m_{h'}=98$~GeV ($m_{h'}=108$~GeV)
and $m_h\simeq 125$~GeV  in both cases.
The dashed lines 
give half of the Higgs masses and the green one the sneutrino mass.}
    \label{fig:massratio}
\end{figure}

The mixing among the Higgs bosons actually imply that the
SM-like Higgs boson also yields a funnel region. This is shown in
\cref{fig:massratio} where we vary the sneutrino mass parameters
but keep the Higgs masses as at the points in \cref{fig:varytanb}
yielding the correct relic density. On the right (left) plot we display
the case for $\tan\beta'=0.972$ ($\tan\beta'=1.032$). For these
values,  we actually have four (two) possible sneutrino masses for the 
smaller (larger) value of $\tan\beta'$
which yield the correct relic density. These correspond to scenarios where 
either the gauge-singlet $h'$ or the SM-like $h$ give rise to the Higgs 
funnel. 

Last but not least we note for completeness that the $Z'$ 
leads to a thermal distribution of the right-handed neutrinos
in the early universe. However, in the preferred keV mass range
for explaining the correct relic density, it is actually a warm dark
matter candidate, which is excluded by observations of the Lyman-$\alpha$ 
forest~\cite{Viel:2013apy,Irsic:2017ixq,Baur:2017stq}. To avoid this, as before, one needs to set a relatively large $M_4$, such that the $\nu_4$ decay quickly. Thus, the right-handed neutrino cannot contribute to the DM in this model.

\section{Conclusions}
\label{sec:conclusions}

In the last years there has been increasing interest
in scenarios where a light right-handed neutrino
could explain the observed relic dark matter
density. This has also been motivated by the fact
that no hint of a WIMP dark matter particle has been
found so far. At the same time there has also been
increasing interest in so-called natural SUSY scenarios 
where the higgsinos are the LSP. While in these scenarios
one can avoid fine-tuning in the Higgs sector, one cannot
explain the observed relic density as the higgsinos 
annihilate very effectively.

We have thus combined
both ideas and considered in a first step a supersymmetric
scenario where the MSSM is extended by right-handed
neutrino superfields where the Yukawa couplings are enhanced by an
inverse seesaw structure. Here we attempted to address
several questions: (i) Is it possible for SUSY loops
to modify the lifetime of the lightest right-handed
neutrino $\nu_4$?  (ii) Is it
possible for an R-sneutrino $\tilde\nu_R$ to be a thermal relic in such
scenarios? (iii) If non-thermal, under which conditions would dark matter production occur?

We have found that the answer to the first two questions is negative
in the more minimal model.
(i) The LHC bounds on the SUSY particles are already
so strong that the corresponding contributions to the
decay $\nu_4 \to \gamma \nu_i$ are suppressed, leading to a change of
the partial widths of at most 10 per-cent with respect to the
SM contribution. 
Consequently, since
the $\nu_4$ should have a mass of a few keV to be an acceptable
DM candidate, we find this scenario to be heavily constrained by cosmological data. (ii) In
scenarios where $\tilde \nu_R$ are the LSP the Yukawa couplings can be at most of 
the order $10^{-5}$. This size is sufficient for the $\tilde \nu_R$
to thermalize, but the corresponding relic density is too large by
many orders of magnitude. The only exception are regions where a sizable
mixing with the $\tilde \nu_L$ is present, as has already
been noticed in the literature. However, we have seen
that these regions lead to charge breaking minima and, thus, are
excluded.

We then find that the only possibility for R-sneutrino dark matter, in the minimal model, is (iii) having a non-thermal FIMP, produced via super-WIMP or freeze-in mechanisms. We find that the observed relic density can be reproduced through both mechanisms, with the super-WIMP depending on the ratio of $\tilde\nu_1$ and NLSP masses. 
The freeze-in is in addition sensitive to $(Y_\nu)_{a4}$. By modifying these parameters, one can make either super-WIMP or freeze-in dominant. We have also found that the NLSP lifetime can be significantly large, which could be in conflict with BBN. 

Although the $\tilde{\nu}_R$ cannot be a thermal DM candidate in the MSSM+$\tilde \nu_R$ framework, as discussed above, it might well be that this is
only an effective model, e.g.\ an additional 
gauge group might be realized in the multi-TeV range. As
an example we have considered the case of an additional
$U(1)_{B-L}$ and extended previous studies by the 
mass hierarchy studied in the MSSM+$\tilde \nu_R$.
The additional $Z'$ in the multi-TeV range implies that 
the $\nu_R$ and $\tilde\nu_R$ get thermalized in the early Universe.
In this class of models a second light Higgs boson
is possible, which mixes somewhat with the SM-like
Higgs boson. We have shown this allows for an explanation of the observed relic density via a Higgs funnel.

\section*{Acknowledgements}

We would like to thank Nicolas Bernal and Carlos Yaguna for discussions.
T.F., Y.L.\ and W.P.\ have been supported by DFG, project nr.\ PO-1337/7-1, 
and by DAAD, project nr.\ 57395885. J.J.P.~acknowledges funding by the 
{\it Direcci\'on de Gesti\'on de la Investigaci\'on} at PUCP, through grant 
DGI-2015-3-0026, and by the DAAD-CONCYTEC project 131-2017-FONDECYT.


\begin{thebibliography}{10}

\bibitem{Papucci:2011wy}
M.~Papucci, J.~T. Ruderman, and A.~Weiler, ``{Natural SUSY Endures},''
  \href{http://dx.doi.org/10.1007/JHEP09(2012)035}{{\em JHEP} {\bfseries 09}
  (2012) 035},
\href{http://arxiv.org/abs/1110.6926}{{\ttfamily arXiv:1110.6926 [hep-ph]}}.

\bibitem{Barducci:2015ffa}
D.~Barducci, A.~Belyaev, A.~K.~M. Bharucha, W.~Porod, and V.~Sanz,
  ``{Uncovering Natural Supersymmetry via the interplay between the LHC and
  Direct Dark Matter Detection},''
  \href{http://dx.doi.org/10.1007/JHEP07(2015)066}{{\em JHEP} {\bfseries 07}
  (2015) 066},
\href{http://arxiv.org/abs/1504.02472}{{\ttfamily arXiv:1504.02472 [hep-ph]}}.

\bibitem{Aprile:2018dbl}
{\bfseries XENON} Collaboration, E.~Aprile {\em et~al.}, ``{Dark Matter Search
  Results from a One Ton-Year Exposure of XENON1T},''
  \href{http://dx.doi.org/10.1103/PhysRevLett.121.111302}{{\em Phys. Rev.
  Lett.} {\bfseries 121} no.~11, (2018) 111302},
\href{http://arxiv.org/abs/1805.12562}{{\ttfamily arXiv:1805.12562
  [astro-ph.CO]}}.

\bibitem{Cirelli:2005uq}
M.~Cirelli, N.~Fornengo, and A.~Strumia, ``{Minimal dark matter},''
  \href{http://dx.doi.org/10.1016/j.nuclphysb.2006.07.012}{{\em Nucl. Phys.}
  {\bfseries B753} (2006) 178--194},
\href{http://arxiv.org/abs/hep-ph/0512090}{{\ttfamily arXiv:hep-ph/0512090
  [hep-ph]}}.

\bibitem{Minkowski:1977sc}
P.~Minkowski, ``{$\mu \to e\gamma$ at a Rate of One Out of $10^{9}$ Muon
  Decays?},''
\href{http://dx.doi.org/10.1016/0370-2693(77)90435-X}{{\em Phys. Lett.}
  {\bfseries B67} (1977) 421--428}.

\bibitem{Yanagida:1979as}
T.~Yanagida, ``{HORIZONTAL SYMMETRY AND MASSES OF NEUTRINOS},''
{\em Conf.Proc.} {\bfseries C7902131} (1979) 95.

\bibitem{Mohapatra:1979ia}
R.~N. Mohapatra and G.~Senjanovic, ``{Neutrino Mass and Spontaneous Parity
  Violation},''
\href{http://dx.doi.org/10.1103/PhysRevLett.44.912}{{\em Phys. Rev. Lett.}
  {\bfseries 44} (1980) 912}.

\bibitem{GellMann:1980vs}
M.~Gell-Mann, P.~Ramond, and R.~Slansky, ``{Complex Spinors and Unified
  Theories},'' {\em Conf. Proc.} {\bfseries C790927} (1979) 315--321,
\href{http://arxiv.org/abs/1306.4669}{{\ttfamily arXiv:1306.4669 [hep-th]}}.

\bibitem{Schechter:1980gr}
J.~Schechter and J.~W.~F. Valle, ``{Neutrino Masses in SU(2) x U(1)
  Theories},''
\href{http://dx.doi.org/10.1103/PhysRevD.22.2227}{{\em Phys. Rev.} {\bfseries
  D22} (1980) 2227}.

\bibitem{Medina:2014bga}
A.~D. Medina, ``{Higgsino-like Dark Matter From Sneutrino Late Decays},''
  \href{http://dx.doi.org/10.1016/j.physletb.2017.04.054}{{\em Phys. Lett.}
  {\bfseries B770} (2017) 161--165},
\href{http://arxiv.org/abs/1409.2560}{{\ttfamily arXiv:1409.2560 [hep-ph]}}.

\bibitem{Hooper:2004dc}
D.~Hooper, J.~March-Russell, and S.~M. West, ``{Asymmetric sneutrino dark
  matter and the Omega(b) / Omega(DM) puzzle},''
  \href{http://dx.doi.org/10.1016/j.physletb.2004.11.047}{{\em Phys. Lett.}
  {\bfseries B605} (2005) 228--236},
\href{http://arxiv.org/abs/hep-ph/0410114}{{\ttfamily arXiv:hep-ph/0410114
  [hep-ph]}}.

\bibitem{Asaka:2005cn}
T.~Asaka, K.~Ishiwata, and T.~Moroi, ``{Right-handed sneutrino as cold dark
  matter},'' \href{http://dx.doi.org/10.1103/PhysRevD.73.051301}{{\em Phys.
  Rev.} {\bfseries D73} (2006) 051301(R)},
\href{http://arxiv.org/abs/hep-ph/0512118}{{\ttfamily arXiv:hep-ph/0512118
  [hep-ph]}}.

\bibitem{Gopalakrishna:2006kr}
S.~Gopalakrishna, A.~de~Gouvea, and W.~Porod, ``{Right-handed sneutrinos as
  nonthermal dark matter},''
  \href{http://dx.doi.org/10.1088/1475-7516/2006/05/005}{{\em JCAP} {\bfseries
  0605} (2006) 005},
\href{http://arxiv.org/abs/hep-ph/0602027}{{\ttfamily arXiv:hep-ph/0602027
  [hep-ph]}}.

\bibitem{Arina:2007tm}
C.~Arina and N.~Fornengo, ``{Sneutrino cold dark matter, a new analysis: Relic
  abundance and detection rates},''
  \href{http://dx.doi.org/10.1088/1126-6708/2007/11/029}{{\em JHEP} {\bfseries
  11} (2007) 029},
\href{http://arxiv.org/abs/0709.4477}{{\ttfamily arXiv:0709.4477 [hep-ph]}}.

\bibitem{Page:2007sh}
V.~Page, ``{Non-thermal right-handed sneutrino dark matter and the
  Omega(DM)/Omega(b) problem},''
  \href{http://dx.doi.org/10.1088/1126-6708/2007/04/021}{{\em JHEP} {\bfseries
  04} (2007) 021},
\href{http://arxiv.org/abs/hep-ph/0701266}{{\ttfamily arXiv:hep-ph/0701266
  [hep-ph]}}.

\bibitem{Belanger:2010cd}
G.~Belanger, M.~Kakizaki, E.~K. Park, S.~Kraml, and A.~Pukhov, ``{Light mixed
  sneutrinos as thermal dark matter},''
  \href{http://dx.doi.org/10.1088/1475-7516/2010/11/017}{{\em JCAP} {\bfseries
  1011} (2010) 017},
\href{http://arxiv.org/abs/1008.0580}{{\ttfamily arXiv:1008.0580 [hep-ph]}}.

\bibitem{Dumont:2012ee}
B.~Dumont, G.~Belanger, S.~Fichet, S.~Kraml, and T.~Schwetz, ``{Mixed sneutrino
  dark matter in light of the 2011 XENON and LHC results},''
  \href{http://dx.doi.org/10.1088/1475-7516/2012/09/013}{{\em JCAP} {\bfseries
  1209} (2012) 013},
\href{http://arxiv.org/abs/1206.1521}{{\ttfamily arXiv:1206.1521 [hep-ph]}}.

\bibitem{DeRomeri:2012qd}
V.~De~Romeri and M.~Hirsch, ``{Sneutrino Dark Matter in Low-scale Seesaw
  Scenarios},'' \href{http://dx.doi.org/10.1007/JHEP12(2012)106}{{\em JHEP}
  {\bfseries 12} (2012) 106},
\href{http://arxiv.org/abs/1209.3891}{{\ttfamily arXiv:1209.3891 [hep-ph]}}.

\bibitem{Banerjee:2016uyt}
S.~Banerjee, G.~Bélanger, B.~Mukhopadhyaya, and P.~D. Serpico, ``{Signatures
  of sneutrino dark matter in an extension of the CMSSM},''
  \href{http://dx.doi.org/10.1007/JHEP07(2016)095}{{\em JHEP} {\bfseries 07}
  (2016) 095},
\href{http://arxiv.org/abs/1603.08834}{{\ttfamily arXiv:1603.08834 [hep-ph]}}.

\bibitem{Choi:2018vdi}
K.-Y. Choi, J.~Kim, and O.~Seto, ``{Thermal production of light Dirac
  right-handed sneutrino dark matter},''
  \href{http://dx.doi.org/10.1016/j.dark.2018.09.006}{{\em Phys. Dark Univ.}
  {\bfseries 22} (2018) 96--100},
\href{http://arxiv.org/abs/1808.00764}{{\ttfamily arXiv:1808.00764 [hep-ph]}}.

\bibitem{Ghosh:2018hly}
A.~Ghosh, T.~Mondal, and B.~Mukhopadhyaya, ``{Right sneutrino with $\Delta L=2$
  masses as nonthermal dark matter},''
  \href{http://dx.doi.org/10.1103/PhysRevD.99.035018}{{\em Phys. Rev.}
  {\bfseries D99} no.~3, (2019) 035018},
\href{http://arxiv.org/abs/1807.04964}{{\ttfamily arXiv:1807.04964 [hep-ph]}}.

\bibitem{Boyarsky:2018tvu}
A.~Boyarsky, M.~Drewes, T.~Lasserre, S.~Mertens, and O.~Ruchayskiy, ``{Sterile
  Neutrino Dark Matter},''
\href{http://arxiv.org/abs/1807.07938}{{\ttfamily arXiv:1807.07938 [hep-ph]}}.

\bibitem{deGouvea:2006wd}
A.~de~Gouvea, S.~Gopalakrishna, and W.~Porod, ``{Stop Decay into Right-handed
  Sneutrino LSP at Hadron Colliders},''
  \href{http://dx.doi.org/10.1088/1126-6708/2006/11/050}{{\em JHEP} {\bfseries
  11} (2006) 050},
\href{http://arxiv.org/abs/hep-ph/0606296}{{\ttfamily arXiv:hep-ph/0606296
  [hep-ph]}}.

\bibitem{Thomas:2007bu}
Z.~Thomas, D.~Tucker-Smith, and N.~Weiner, ``{Mixed Sneutrinos, Dark Matter and
  the CERN LHC},'' \href{http://dx.doi.org/10.1103/PhysRevD.77.115015}{{\em
  Phys. Rev.} {\bfseries D77} (2008) 115015},
\href{http://arxiv.org/abs/0712.4146}{{\ttfamily arXiv:0712.4146 [hep-ph]}}.

\bibitem{Choudhury:2008gb}
D.~Choudhury, S.~K. Gupta, and B.~Mukhopadhyaya, ``{Right sneutrinos in a
  supergravity model and the signals of a stable stop at the Large Hadron
  Collider},'' \href{http://dx.doi.org/10.1103/PhysRevD.78.015023}{{\em Phys.
  Rev.} {\bfseries D78} (2008) 015023},
\href{http://arxiv.org/abs/0804.3560}{{\ttfamily arXiv:0804.3560 [hep-ph]}}.

\bibitem{Cerna-Velazco:2017cmn}
N.~Cerna-Velazco, T.~Faber, J.~Jones-Perez, and W.~Porod, ``{Constraining
  sleptons at the LHC in a supersymmetric low-scale seesaw scenario},''
  \href{http://dx.doi.org/10.1140/epjc/s10052-017-5231-9}{{\em Eur. Phys. J.}
  {\bfseries C77} no.~10, (2017) 661},
\href{http://arxiv.org/abs/1705.06583}{{\ttfamily arXiv:1705.06583 [hep-ph]}}.

\bibitem{Chatterjee:2017nyx}
A.~Chatterjee, J.~Dutta, and S.~K. Rai, ``{Natural SUSY at LHC with
  Right-Sneutrino LSP},'' \href{http://dx.doi.org/10.1007/JHEP06(2018)042}{{\em
  JHEP} {\bfseries 06} (2018) 042},
\href{http://arxiv.org/abs/1710.10617}{{\ttfamily arXiv:1710.10617 [hep-ph]}}.

\bibitem{Banerjee:2018uut}
S.~Banerjee, G.~Bélanger, A.~Ghosh, and B.~Mukhopadhyaya, ``{Long-lived stau,
  sneutrino dark matter and right-slepton spectrum},''
  \href{http://dx.doi.org/10.1007/JHEP09(2018)143}{{\em JHEP} {\bfseries 09}
  (2018) 143},
\href{http://arxiv.org/abs/1806.04488}{{\ttfamily arXiv:1806.04488 [hep-ph]}}.

\bibitem{Casas:2001sr}
J.~A. Casas and A.~Ibarra, ``{Oscillating neutrinos and $\mu \to e \gamma$},''
  \href{http://dx.doi.org/10.1016/S0550-3213(01)00475-8}{{\em Nucl. Phys.}
  {\bfseries B618} (2001) 171--204},
\href{http://arxiv.org/abs/hep-ph/0103065}{{\ttfamily arXiv:hep-ph/0103065
  [hep-ph]}}.

\bibitem{Donini:2012tt}
A.~Donini, P.~Hernandez, J.~Lopez-Pavon, M.~Maltoni, and T.~Schwetz, ``{The
  minimal 3+2 neutrino model versus oscillation anomalies},''
  \href{http://dx.doi.org/10.1007/JHEP07(2012)161}{{\em JHEP} {\bfseries 07}
  (2012) 161},
\href{http://arxiv.org/abs/1205.5230}{{\ttfamily arXiv:1205.5230 [hep-ph]}}.

\bibitem{Ibarra:2010xw}
A.~Ibarra, E.~Molinaro, and S.~T. Petcov, ``{TeV Scale See-Saw Mechanisms of
  Neutrino Mass Generation, the Majorana Nature of the Heavy Singlet Neutrinos
  and $(\beta\beta)_{0\nu}$-Decay},''
  \href{http://dx.doi.org/10.1007/JHEP09(2010)108}{{\em JHEP} {\bfseries 09}
  (2010) 108},
\href{http://arxiv.org/abs/1007.2378}{{\ttfamily arXiv:1007.2378 [hep-ph]}}.

\bibitem{LopezPavon:2012zg}
J.~Lopez-Pavon, S.~Pascoli, and C.-f. Wong, ``{Can heavy neutrinos dominate
  neutrinoless double beta decay?},''
  \href{http://dx.doi.org/10.1103/PhysRevD.87.093007}{{\em Phys. Rev.}
  {\bfseries D87} no.~9, (2013) 093007},
\href{http://arxiv.org/abs/1209.5342}{{\ttfamily arXiv:1209.5342 [hep-ph]}}.

\bibitem{Gago:2015vma}
A.~M. Gago, P.~Hernández, J.~Jones-Pérez, M.~Losada, and A.~Moreno~Briceño,
  ``{Probing the Type I Seesaw Mechanism with Displaced Vertices at the LHC},''
  \href{http://dx.doi.org/10.1140/epjc/s10052-015-3693-1}{{\em Eur. Phys. J.}
  {\bfseries C75} no.~10, (2015) 470},
\href{http://arxiv.org/abs/1505.05880}{{\ttfamily arXiv:1505.05880 [hep-ph]}}.

\bibitem{Hernandez:2018cgc}
P.~Hernández, J.~Jones-Pérez, and O.~Suárez-Navarro, ``{Majorana vs
  Pseudo-Dirac Neutrinos at the ILC},''
\href{http://arxiv.org/abs/1810.07210}{{\ttfamily arXiv:1810.07210 [hep-ph]}}.

\bibitem{deSalas:2017kay}
P.~F. de~Salas, D.~V. Forero, C.~A. Ternes, M.~Tortola, and J.~W.~F. Valle,
  ``{Status of neutrino oscillations 2018: 3$\sigma$ hint for normal mass
  ordering and improved CP sensitivity},''
  \href{http://dx.doi.org/10.1016/j.physletb.2018.06.019}{{\em Phys. Lett.}
  {\bfseries B782} (2018) 633--640},
\href{http://arxiv.org/abs/1708.01186}{{\ttfamily arXiv:1708.01186 [hep-ph]}}.

\bibitem{Esteban:2018azc}
I.~Esteban, M.~C. Gonzalez-Garcia, A.~Hernandez-Cabezudo, M.~Maltoni, and
  T.~Schwetz, ``{Global analysis of three-flavour neutrino oscillations:
  synergies and tensions in the determination of $\theta_23, \delta_CP$, and
  the mass ordering},''
\href{http://arxiv.org/abs/1811.05487}{{\ttfamily arXiv:1811.05487 [hep-ph]}}.

\bibitem{Tanabashi:2018oca}
{\bfseries Particle Data Group} Collaboration, M.~Tanabashi {\em et~al.},
  ``{Review of Particle Physics},''
\href{http://dx.doi.org/10.1103/PhysRevD.98.030001}{{\em Phys. Rev.} {\bfseries
  D98} no.~3, (2018) 030001}.

\bibitem{Dodelson:1993je}
S.~Dodelson and L.~M. Widrow, ``{Sterile-neutrinos as dark matter},''
  \href{http://dx.doi.org/10.1103/PhysRevLett.72.17}{{\em Phys. Rev. Lett.}
  {\bfseries 72} (1994) 17--20},
\href{http://arxiv.org/abs/hep-ph/9303287}{{\ttfamily arXiv:hep-ph/9303287
  [hep-ph]}}.

\bibitem{Shi:1998km}
X.-D. Shi and G.~M. Fuller, ``{A New dark matter candidate: Nonthermal sterile
  neutrinos},'' \href{http://dx.doi.org/10.1103/PhysRevLett.82.2832}{{\em Phys.
  Rev. Lett.} {\bfseries 82} (1999) 2832--2835},
\href{http://arxiv.org/abs/astro-ph/9810076}{{\ttfamily arXiv:astro-ph/9810076
  [astro-ph]}}.

\bibitem{Ng:2019gch}
K.~C.~Y. Ng, B.~M. Roach, K.~Perez, J.~F. Beacom, S.~Horiuchi, R.~Krivonos, and
  D.~R. Wik, ``{New Constraints on Sterile Neutrino Dark Matter from $NuSTAR$
  M31 Observations},''
\href{http://arxiv.org/abs/1901.01262}{{\ttfamily arXiv:1901.01262
  [astro-ph.HE]}}.

\bibitem{Drewes:2019mhg}
M.~Drewes, ``{On the Minimal Mixing of Heavy Neutrinos},''
\href{http://arxiv.org/abs/1904.11959}{{\ttfamily arXiv:1904.11959 [hep-ph]}}.

\bibitem{Haber:1988px}
H.~E. Haber and D.~Wyler, ``{RADIATIVE NEUTRALINO DECAY},''
\href{http://dx.doi.org/10.1016/0550-3213(89)90143-0}{{\em Nucl. Phys.}
  {\bfseries B323} (1989) 267--310}.

\bibitem{Vincent:2014rja}
A.~C. Vincent, E.~F. Martinez, P.~Hernández, M.~Lattanzi, and O.~Mena,
  ``{Revisiting cosmological bounds on sterile neutrinos},''
  \href{http://dx.doi.org/10.1088/1475-7516/2015/04/006}{{\em JCAP} {\bfseries
  1504} no.~04, (2015) 006},
\href{http://arxiv.org/abs/1408.1956}{{\ttfamily arXiv:1408.1956
  [astro-ph.CO]}}.

\bibitem{Shakya:2016oxf}
B.~Shakya and J.~D. Wells, ``{Sterile Neutrino Dark Matter with
  Supersymmetry},'' \href{http://dx.doi.org/10.1103/PhysRevD.96.031702}{{\em
  Phys. Rev.} {\bfseries D96} no.~3, (2017) 031702},
\href{http://arxiv.org/abs/1611.01517}{{\ttfamily arXiv:1611.01517 [hep-ph]}}.

\bibitem{Bertone:2004pz}
G.~Bertone, D.~Hooper, and J.~Silk, ``{Particle dark matter: Evidence,
  candidates and constraints},''
  \href{http://dx.doi.org/10.1016/j.physrep.2004.08.031}{{\em Phys. Rept.}
  {\bfseries 405} (2005) 279--390},
\href{http://arxiv.org/abs/hep-ph/0404175}{{\ttfamily arXiv:hep-ph/0404175
  [hep-ph]}}.

\bibitem{Feng:2003xh}
J.~L. Feng, A.~Rajaraman, and F.~Takayama, ``{Superweakly interacting massive
  particles},'' \href{http://dx.doi.org/10.1103/PhysRevLett.91.011302}{{\em
  Phys. Rev. Lett.} {\bfseries 91} (2003) 011302},
\href{http://arxiv.org/abs/hep-ph/0302215}{{\ttfamily arXiv:hep-ph/0302215
  [hep-ph]}}.

\bibitem{Feng:2003uy}
J.~L. Feng, A.~Rajaraman, and F.~Takayama, ``{SuperWIMP dark matter signals
  from the early universe},''
  \href{http://dx.doi.org/10.1103/PhysRevD.68.063504}{{\em Phys. Rev.}
  {\bfseries D68} (2003) 063504},
\href{http://arxiv.org/abs/hep-ph/0306024}{{\ttfamily arXiv:hep-ph/0306024
  [hep-ph]}}.

\bibitem{Asaka:2006fs}
T.~Asaka, K.~Ishiwata, and T.~Moroi, ``{Right-handed sneutrino as cold dark
  matter of the universe},''
  \href{http://dx.doi.org/10.1103/PhysRevD.75.065001}{{\em Phys. Rev.}
  {\bfseries D75} (2007) 065001},
\href{http://arxiv.org/abs/hep-ph/0612211}{{\ttfamily arXiv:hep-ph/0612211
  [hep-ph]}}.

\bibitem{Hall:2009bx}
L.~J. Hall, K.~Jedamzik, J.~March-Russell, and S.~M. West, ``{Freeze-In
  Production of FIMP Dark Matter},''
  \href{http://dx.doi.org/10.1007/JHEP03(2010)080}{{\em JHEP} {\bfseries 03}
  (2010) 080},
\href{http://arxiv.org/abs/0911.1120}{{\ttfamily arXiv:0911.1120 [hep-ph]}}.

\bibitem{Belanger:2018sti}
G.~Bélanger {\em et~al.}, ``{LHC-friendly minimal freeze-in models},''
\href{http://arxiv.org/abs/1811.05478}{{\ttfamily arXiv:1811.05478 [hep-ph]}}.

\bibitem{Staub:2008uz}
F.~Staub, ``{SARAH},''
\href{http://arxiv.org/abs/0806.0538}{{\ttfamily arXiv:0806.0538 [hep-ph]}}.

\bibitem{Staub:2013tta}
F.~Staub, ``{SARAH 4 : A tool for (not only SUSY) model builders},''
  \href{http://dx.doi.org/10.1016/j.cpc.2014.02.018}{{\em Comput. Phys.
  Commun.} {\bfseries 185} (2014) 1773--1790},
\href{http://arxiv.org/abs/1309.7223}{{\ttfamily arXiv:1309.7223 [hep-ph]}}.

\bibitem{Staub:2012pb}
F.~Staub, ``{SARAH 3.2: Dirac Gauginos, UFO output, and more},''
  \href{http://dx.doi.org/10.1016/j.cpc.2013.02.019}{{\em Comput. Phys.
  Commun.} {\bfseries 184} (2013) 1792--1809},
\href{http://arxiv.org/abs/1207.0906}{{\ttfamily arXiv:1207.0906 [hep-ph]}}.

\bibitem{Staub:2010jh}
F.~Staub, ``{Automatic Calculation of supersymmetric Renormalization Group
  Equations and Self Energies},''
  \href{http://dx.doi.org/10.1016/j.cpc.2010.11.030}{{\em Comput. Phys.
  Commun.} {\bfseries 182} (2011) 808--833},
\href{http://arxiv.org/abs/1002.0840}{{\ttfamily arXiv:1002.0840 [hep-ph]}}.

\bibitem{Staub:2009bi}
F.~Staub, ``{From Superpotential to Model Files for FeynArts and
  CalcHep/CompHep},'' \href{http://dx.doi.org/10.1016/j.cpc.2010.01.011}{{\em
  Comput. Phys. Commun.} {\bfseries 181} (2010) 1077--1086},
\href{http://arxiv.org/abs/0909.2863}{{\ttfamily arXiv:0909.2863 [hep-ph]}}.

\bibitem{Porod:2003um}
W.~Porod, ``{SPheno, a program for calculating supersymmetric spectra, SUSY
  particle decays and SUSY particle production at e+ e- colliders},''
  \href{http://dx.doi.org/10.1016/S0010-4655(03)00222-4}{{\em Comput. Phys.
  Commun.} {\bfseries 153} (2003) 275--315},
\href{http://arxiv.org/abs/hep-ph/0301101}{{\ttfamily arXiv:hep-ph/0301101
  [hep-ph]}}.

\bibitem{Porod:2011nf}
W.~Porod and F.~Staub, ``{SPheno 3.1: Extensions including flavour, CP-phases
  and models beyond the MSSM},''
  \href{http://dx.doi.org/10.1016/j.cpc.2012.05.021}{{\em Comput. Phys.
  Commun.} {\bfseries 183} (2012) 2458--2469},
\href{http://arxiv.org/abs/1104.1573}{{\ttfamily arXiv:1104.1573 [hep-ph]}}.

\bibitem{Staub:2011dp}
F.~Staub, T.~Ohl, W.~Porod, and C.~Speckner, ``{A Tool Box for Implementing
  Supersymmetric Models},''
  \href{http://dx.doi.org/10.1016/j.cpc.2012.04.013}{{\em Comput. Phys.
  Commun.} {\bfseries 183} (2012) 2165--2206},
\href{http://arxiv.org/abs/1109.5147}{{\ttfamily arXiv:1109.5147 [hep-ph]}}.

\bibitem{Belanger:2018mqt}
G.~Bélanger, F.~Boudjema, A.~Goudelis, A.~Pukhov, and B.~Zaldivar,
  ``{micrOMEGAs5.0 : Freeze-in},''
  \href{http://dx.doi.org/10.1016/j.cpc.2018.04.027}{{\em Comput. Phys.
  Commun.} {\bfseries 231} (2018) 173--186},
\href{http://arxiv.org/abs/1801.03509}{{\ttfamily arXiv:1801.03509 [hep-ph]}}.

\bibitem{Garny:2018ali}
M.~Garny and J.~Heisig, ``{Interplay of super-WIMP and freeze-in production of
  dark matter},'' \href{http://dx.doi.org/10.1103/PhysRevD.98.095031}{{\em
  Phys. Rev.} {\bfseries D98} no.~9, (2018) 095031},
\href{http://arxiv.org/abs/1809.10135}{{\ttfamily arXiv:1809.10135 [hep-ph]}}.

\bibitem{Asaka:2006ek}
T.~Asaka, M.~Shaposhnikov, and A.~Kusenko, ``{Opening a new window for warm
  dark matter},'' \href{http://dx.doi.org/10.1016/j.physletb.2006.05.067}{{\em
  Phys. Lett.} {\bfseries B638} (2006) 401--406},
\href{http://arxiv.org/abs/hep-ph/0602150}{{\ttfamily arXiv:hep-ph/0602150
  [hep-ph]}}.

\bibitem{Kawasaki:2004qu}
M.~Kawasaki, K.~Kohri, and T.~Moroi, ``{Big-Bang nucleosynthesis and hadronic
  decay of long-lived massive particles},''
  \href{http://dx.doi.org/10.1103/PhysRevD.71.083502}{{\em Phys. Rev.}
  {\bfseries D71} (2005) 083502},
\href{http://arxiv.org/abs/astro-ph/0408426}{{\ttfamily arXiv:astro-ph/0408426
  [astro-ph]}}.

\bibitem{Ishiwata:2009gs}
K.~Ishiwata, M.~Kawasaki, K.~Kohri, and T.~Moroi, ``{Right-handed sneutrino
  dark matter and big-bang nucleosynthesis},''
  \href{http://dx.doi.org/10.1016/j.physletb.2010.04.054}{{\em Phys. Lett.}
  {\bfseries B689} (2010) 163--168},
\href{http://arxiv.org/abs/0912.0781}{{\ttfamily arXiv:0912.0781 [hep-ph]}}.

\bibitem{Kawasaki:2017bqm}
M.~Kawasaki, K.~Kohri, T.~Moroi, and Y.~Takaesu, ``{Revisiting Big-Bang
  Nucleosynthesis Constraints on Long-Lived Decaying Particles},''
  \href{http://dx.doi.org/10.1103/PhysRevD.97.023502}{{\em Phys. Rev.}
  {\bfseries D97} no.~2, (2018) 023502},
\href{http://arxiv.org/abs/1709.01211}{{\ttfamily arXiv:1709.01211 [hep-ph]}}.

\bibitem{Aghanim:2018eyx}
{\bfseries Planck} Collaboration, N.~Aghanim {\em et~al.}, ``{Planck 2018
  results. VI. Cosmological parameters},''
\href{http://arxiv.org/abs/1807.06209}{{\ttfamily arXiv:1807.06209
  [astro-ph.CO]}}.

\bibitem{ArkaniHamed:2000bq}
N.~Arkani-Hamed, L.~J. Hall, H.~Murayama, D.~Smith, and N.~Weiner,
  ``{Small neutrino masses from supersymmetry breaking},''
  \href{http://dx.doi.org/10.1103/PhysRevD.64.115011}{{\em Phys. Rev.}
  {\bfseries D64} (2001) 115011},
\href{http://arxiv.org/abs/hep-ph/0006312}{{\ttfamily arXiv:hep-ph/0006312
  [hep-ph]}}.

\bibitem{Casas:1995pd}
J.~A. Casas, A.~Lleyda, and C.~Munoz, ``{Strong constraints on the parameter
  space of the MSSM from charge and color breaking minima},''
  \href{http://dx.doi.org/10.1016/0550-3213(96)00194-0}{{\em Nucl. Phys.}
  {\bfseries B471} (1996) 3--58},
\href{http://arxiv.org/abs/hep-ph/9507294}{{\ttfamily arXiv:hep-ph/9507294
  [hep-ph]}}.

\bibitem{Kakizaki:2015nua}
M.~Kakizaki, E.-K. Park, J.-h. Park, and A.~Santa, ``{Phenomenological
  constraints on light mixed sneutrino dark matter scenarios},''
  \href{http://dx.doi.org/10.1016/j.physletb.2015.07.030}{{\em Phys. Lett.}
  {\bfseries B749} (2015) 44--49},
\href{http://arxiv.org/abs/1503.06783}{{\ttfamily arXiv:1503.06783 [hep-ph]}}.

\bibitem{Camargo-Molina:2013sta}
J.~E. Camargo-Molina, B.~O'Leary, W.~Porod, and F.~Staub, ``{Stability of the
  CMSSM against sfermion VEVs},''
  \href{http://dx.doi.org/10.1007/JHEP12(2013)103}{{\em JHEP} {\bfseries 12}
  (2013) 103},
\href{http://arxiv.org/abs/1309.7212}{{\ttfamily arXiv:1309.7212 [hep-ph]}}.

\bibitem{CamargoMolina:2012hv}
J.~E. Camargo-Molina, B.~O'Leary, W.~Porod, and F.~Staub, ``{The Stability Of
  R-Parity In Supersymmetric Models Extended By $U(1)_{B-L}$},''
  \href{http://dx.doi.org/10.1103/PhysRevD.88.015033}{{\em Phys. Rev.}
  {\bfseries D88} (2013) 015033},
\href{http://arxiv.org/abs/1212.4146}{{\ttfamily arXiv:1212.4146 [hep-ph]}}.

\bibitem{Camargo-Molina:2013qva}
J.~E. Camargo-Molina, B.~O'Leary, W.~Porod, and F.~Staub,
  ``{$\mathbf{Vevacious}$: A Tool For Finding The Global Minima Of One-Loop
  Effective Potentials With Many Scalars},''
  \href{http://dx.doi.org/10.1140/epjc/s10052-013-2588-2}{{\em Eur. Phys. J.}
  {\bfseries C73} no.~10, (2013) 2588},
\href{http://arxiv.org/abs/1307.1477}{{\ttfamily arXiv:1307.1477 [hep-ph]}}.

\bibitem{Lee:2007mt}
H.-S. Lee, K.~T. Matchev, and S.~Nasri, ``{Revival of the thermal sneutrino
  dark matter},'' \href{http://dx.doi.org/10.1103/PhysRevD.76.041302}{{\em
  Phys. Rev.} {\bfseries D76} (2007) 041302(R)},
\href{http://arxiv.org/abs/hep-ph/0702223}{{\ttfamily arXiv:hep-ph/0702223
  [HEP-PH]}}.

\bibitem{Belanger:2011rs}
G.~Belanger, J.~Da~Silva, and A.~Pukhov, ``{The Right-handed sneutrino as
  thermal dark matter in U(1) extensions of the MSSM},''
  \href{http://dx.doi.org/10.1088/1475-7516/2011/12/014}{{\em JCAP} {\bfseries
  1112} (2011) 014},
\href{http://arxiv.org/abs/1110.2414}{{\ttfamily arXiv:1110.2414 [hep-ph]}}.

\bibitem{Bandyopadhyay:2011qm}
P.~Bandyopadhyay, E.~J. Chun, and J.-C. Park, ``{Right-handed sneutrino dark
  matter in $\mathbf{U(1)'}$ seesaw models and its signatures at the LHC},''
  \href{http://dx.doi.org/10.1007/JHEP06(2011)129}{{\em JHEP} {\bfseries 06}
  (2011) 129},
\href{http://arxiv.org/abs/1105.1652}{{\ttfamily arXiv:1105.1652 [hep-ph]}}.

\bibitem{Basso:2012gz}
L.~Basso, B.~O'Leary, W.~Porod, and F.~Staub, ``{Dark matter scenarios in the
  minimal SUSY B-L model},''
  \href{http://dx.doi.org/10.1007/JHEP09(2012)054}{{\em JHEP} {\bfseries 09}
  (2012) 054},
\href{http://arxiv.org/abs/1207.0507}{{\ttfamily arXiv:1207.0507 [hep-ph]}}.

\bibitem{Frank:2017tsm}
M.~Frank, B.~Fuks, K.~Huitu, S.~K. Rai, and H.~Waltari, ``{Resonant slepton
  production and right sneutrino dark matter in left-right supersymmetry},''
  \href{http://dx.doi.org/10.1007/JHEP05(2017)015}{{\em JHEP} {\bfseries 05}
  (2017) 015},
\href{http://arxiv.org/abs/1702.02112}{{\ttfamily arXiv:1702.02112 [hep-ph]}}.

\bibitem{Cerdeno:2009dv}
D.~G. Cerdeno and O.~Seto, ``{Right-handed sneutrino dark matter in the
  NMSSM},'' \href{http://dx.doi.org/10.1088/1475-7516/2009/08/032}{{\em JCAP}
  {\bfseries 0908} (2009) 032},
\href{http://arxiv.org/abs/0903.4677}{{\ttfamily arXiv:0903.4677 [hep-ph]}}.

\bibitem{Cerdeno:2011qv}
D.~G. Cerdeno, J.-H. Huh, M.~Peiro, and O.~Seto, ``{Very light right-handed
  sneutrino dark matter in the NMSSM},''
  \href{http://dx.doi.org/10.1088/1475-7516/2011/11/027}{{\em JCAP} {\bfseries
  1111} (2011) 027},
\href{http://arxiv.org/abs/1108.0978}{{\ttfamily arXiv:1108.0978 [hep-ph]}}.

\bibitem{Deppisch:2008bp}
F.~Deppisch and A.~Pilaftsis, ``{Thermal Right-Handed Sneutrino Dark Matter in
  the F(D)-Term Model of Hybrid Inflation},''
  \href{http://dx.doi.org/10.1088/1126-6708/2008/10/080}{{\em JHEP} {\bfseries
  10} (2008) 080},
\href{http://arxiv.org/abs/0808.0490}{{\ttfamily arXiv:0808.0490 [hep-ph]}}.

\bibitem{OLeary:2011vlq}
B.~O'Leary, W.~Porod, and F.~Staub, ``{Mass spectrum of the minimal SUSY B-L
  model},'' \href{http://dx.doi.org/10.1007/JHEP05(2012)042}{{\em JHEP}
  {\bfseries 05} (2012) 042},
\href{http://arxiv.org/abs/1112.4600}{{\ttfamily arXiv:1112.4600 [hep-ph]}}.

\bibitem{Barger:2008wn}
V.~Barger, P.~Fileviez~Perez, and S.~Spinner, ``{Minimal gauged $U(1)_{(B-L)}$
  model with spontaneous R-parity violation},''
  \href{http://dx.doi.org/10.1103/PhysRevLett.102.181802}{{\em Phys. Rev.
  Lett.} {\bfseries 102} (2009) 181802},
\href{http://arxiv.org/abs/0812.3661}{{\ttfamily arXiv:0812.3661 [hep-ph]}}.

\bibitem{Bechtle:2013wla}
P.~Bechtle, O.~Brein, S.~Heinemeyer, O.~Stål, T.~Stefaniak, G.~Weiglein, and
  K.~E. Williams, ``{$\mathsf{HiggsBounds}-4$: Improved Tests of Extended Higgs
  Sectors against Exclusion Bounds from LEP, the Tevatron and the LHC},''
  \href{http://dx.doi.org/10.1140/epjc/s10052-013-2693-2}{{\em Eur. Phys. J.}
  {\bfseries C74} no.~3, (2014) 2693},
\href{http://arxiv.org/abs/1311.0055}{{\ttfamily arXiv:1311.0055 [hep-ph]}}.

\bibitem{Viel:2013apy}
M.~Viel, G.~D. Becker, J.~S. Bolton, and M.~G. Haehnelt, ``{Warm dark matter as
  a solution to the small scale crisis: New constraints from high redshift
  Lyman-$\alpha$ forest data},''
  \href{http://dx.doi.org/10.1103/PhysRevD.88.043502}{{\em Phys. Rev.}
  {\bfseries D88} (2013) 043502},
\href{http://arxiv.org/abs/1306.2314}{{\ttfamily arXiv:1306.2314
  [astro-ph.CO]}}.

\bibitem{Irsic:2017ixq}
V.~Iršič {\em et~al.}, ``{New Constraints on the free-streaming of warm dark
  matter from intermediate and small scale Lyman-$\alpha$ forest data},''
  \href{http://dx.doi.org/10.1103/PhysRevD.96.023522}{{\em Phys. Rev.}
  {\bfseries D96} no.~2, (2017) 023522},
\href{http://arxiv.org/abs/1702.01764}{{\ttfamily arXiv:1702.01764
  [astro-ph.CO]}}.

\bibitem{Baur:2017stq}
J.~Baur, N.~Palanque-Delabrouille, C.~Yeche, A.~Boyarsky, O.~Ruchayskiy,
  {\'E}.~Armengaud, and J.~Lesgourgues, ``{Constraints from Ly-$\alpha$ forests
  on non-thermal dark matter including resonantly-produced sterile
  neutrinos},'' \href{http://dx.doi.org/10.1088/1475-7516/2017/12/013}{{\em
  JCAP} {\bfseries 1712} no.~12, (2017) 013},
\href{http://arxiv.org/abs/1706.03118}{{\ttfamily arXiv:1706.03118
  [astro-ph.CO]}}.

\end{thebibliography}
\end{document}